\documentclass[preprint,12pt, 2p, times]{elsarticle}

\usepackage{amssymb,amsmath}

\usepackage{array}
\usepackage{color}
\usepackage{mathdots}
\usepackage{graphicx} % Required for including pictures
\usepackage{wrapfig} % Allows in-line images
\usepackage{aastex_hack}

\journal{Journal of Computational Physics}

\begin{document}

\begin{frontmatter}



\title{A Six-moment Multi-fluid Plasma Model}


\author{Zhenguang Huang, G\'abor T\'oth, Bart van der Holst, Yuxi Chen, Tamas Gombosi}

\address{Climate and Space Sciences and Engineering, University of Michigan, Ann Arbor, MI 48109, USA}

\begin{abstract}

We present a six-moment multi-fluid model, which solves the governing equations for both ions
and electrons, with pressure anisotropy along and perpendicular to the magnetic field direction,
as well as the complete set of Maxwell equations. This set of equations includes the Hall effect, 
different temperatures for different species and pressure anisotropy. It is more comprehensive than
the five-moment equations with isotropic pressures and significantly 
less expensive than the ten-moment equations with a full pressure tensors. 
Similarly to the five- and ten-moment equations, 
the wave speeds are naturally limited by the speed of light, which eliminates the issue of
unlimited whistler wave speeds present in Hall magnetohydrodynamics (MHD).
It is also possible to simulate multiple negatively charged fluids, 
which cannot be done in MHD models. The six-moment model is a reasonable description of the plasma 
outside magnetic reconnection regions and therefore well-suited to be coupled
with an embedded particle-in-cell model that covers the reconnection region. 
Our numerical implementation
uses a point-implicit scheme for the stiff source terms, and we use a second-order accurate
Rusanov-type scheme with carefully selected wave speeds. For the plasma variables and the 
magnetic field the maximum wave speed is based on the fast
magnetosonic speed of MHD with anisotropic pressures that we derive. 
For the electric field related variables the speed of light is used. 
The divergence of the magnetic field and Gauss's law are controlled with a hyperbolic-parabolic
scheme. We present a number of numerical tests to demonstrate that this numerical model
is robust without being excessively diffusive.

\end{abstract}


\end{frontmatter}

\section{Introduction}
Magnetohydrodynamics (MHD) simulations have been widely carried out to understand the mechanisms behind
different phenomena in plasma physics. 
MHD models assume that the Larmor radius (gyro radius) is much smaller
than the characteristic length scale and the particle distribution function can be described by the fluid equations (continuity, momentum and pressure/energy equations). Magnetic ($\mathbf{B}$) and electric ($\mathbf{E}$) fields are needed to solve the
governing equations for the ions and electrons. In the MHD approximation, 
the mass of the electrons is neglected so that we can obtain the approximate electric field
from the electron momentum equation. In such an approximation, the magnetic field is frozen into the electron fluid.
Further simplifications include ignoring the velocity difference between the electrons and ions in the induction equation and assuming equal temperatures and Maxwellian distributions for both electrons and ions, which leads to the ideal MHD approximation. 
In ideal MHD, the magnetic field lines are frozen into the plasma consisting of co-moving ions and electrons.
The frozen-in condition can be relaxed by taking into account resistivity, however, resistivity is negligible for collisionless plasmas found in space and astrophysics, for example. 
Another improvement is to include the velocity difference between
ions and electrons in the magnetic induction equation. The resulting Hall MHD model includes some of the ion physics. 
A further step towards a kinetic description is to allow for different electron and ion pressures and allow for pressure anisotropy (for example \cite{Gombosi_1991, Meng_2012}). 

The MHD (including ideal, resistive, Hall and MHD with anisotropic pressure) description neglects electron inertia due to the finite mass of electrons and assumes perfect charge neutrality. The five-moment equations of Shumlak and Loverich \cite{Shumlak_2003} 
remove the assumption of the massless electrons and solve the full set of Maxwell equations to obtain 
the electric and magnetic fields. In addition, the full set of hydrodynamic equations with the Lorentz force 
on the right-hand-side are solved separately for the electron and ion fluids, so that the electron mass is taken into account and charge separation is allowed. 
Further development \cite{Hakim_2008, Wang_2015, Wang_2018arXiv} lead to the ten-moment two-fluid
plasma model that solves for full pressure tensors for both the 
electron and ion fluids. Wang et al. \cite{Wang_2015} compare their five- and ten-moment two-fluid plasma models with a Particle-In-Cell (PIC) kinetic model and show that their five- and ten-moment models can reproduce many important kinetic features observed in the PIC simulation.
Alvarez-Laguna et al. \cite{Alvarez-Laguna_2018} recently proposed a new numerical method, which contains implicit time integration to handle the stiffness of the system and properly scaling of the numerical dissipation from the electromagnetic field solver to the plasma flow solver, to simulate the multi-fluid plasma system.

In this manuscript, we propose another approach, a six-moment multi-fluid plasma model, which is in-between the five- and ten-moment models, by introducing pressure anisotropy for both ions and electrons along and perpendicular to the magnetic field direction. The six-moment approximation requires only two pressure components (parallel and perpendicular) per fluid instead of the six independent components of the full pressure tensors, and the six-moment equations are significantly simpler than the ten-moment equations. This means that the six-moment equations are less expensive to solve. In addition, the six-moment equations are likely to be valid in the vast majority of the plasma system where the electrons and ions are both magnetized, so the off-diagonal terms of the pressure tensor are negligible. Near reconnection regions the six-moment approximation is not valid, but even the ten-moment approximation has a difficult time to reproduce all aspects of kinetic reconnection, although there has been some promising progress \cite{Wang_2015}. An alternative approach, that we plan to employ in the future,  is to use an embedded particle-in-cell (PIC) model \cite{Daldorff_2014, Chen_2018} to cover the reconnection region. We expect the six-moment equations to provide a good fluid model that can be coupled effectively to the embedded PIC model covering the reconnection site. We expect that using the six-moment model allows reducing the size of the PIC domain compared to the case when the fluid model is
simpler (MHD or Hall MHD). An additional feature of the six-moment (also true for the five- and ten-moment) equations is that one can allow for multiple electron fluids or a mixture of electrons and negatively charged ions. These situations cannot be handled with the usual MHD models, as the densities and velocities of the multiple electron fluids cannot be determined from the 
charge neutrality and electric current. 

In the following section we present the six-moment equations, then we derive the characteristic speeds 
in section \ref{sec:wavespeed} that are used in the discretization. 
The numerical scheme employing a reduced numerical dissipation in combination with a point-implicit discretization of the stiff source terms is discussed in section~\ref{sec:discretization}. 
We present several numerical tests in section~\ref{sec:tests} to demonstrate
the capabilities of our six-moment model, and conclude with section \ref{sec:conclusion}.

\section{Model Equations}
\label{sec:model}

The six-moment equations are an extension of the five-moment equations \citep{Shumlak_2003, Hakim_2006} by introducing pressure anisotropy for both ions and electrons
\cite{Gombosi_1991, Meng_2012}. 
Under this assumption, the pressure tensor can be approximated with
$\mathbf{P} = p_{\perp} \mathbf{I} + (p_{\parallel} - p_{\perp}) \mathbf{bb}$, 
where $\mathbf{I}$ is the identity matrix,  $\mathbf{b}$ is the unit vector along the magnetic field direction, $p_{\parallel}$ is the pressure along the parallel direction of the magnetic
field and $p_{\perp}$ is the pressure in the perpendicular direction.
For monatomic gases, 
the six-moment equations for all charged fluids (indexed by $s$) can be written as:
%{\setlength{\mathindent}{\parindent}{
\begin{subequations}
\label{eqn:sixmoment}
\begin{align}
& \frac{\partial \rho_s}{\partial t} + \nabla \cdot (\rho_s \mathbf{u_s}) = 0	\\
& \frac{\partial \rho_s \mathbf{u_s}}{\partial t}  + \nabla \cdot 
   \left[\rho_s \mathbf{u_s}\mathbf{u_s} + p_{s\perp} \mathbf{I} 
         + (p_{s\parallel} - p_{s\perp})\mathbf{bb}\right] =
  \frac{q_s}{m_s}\rho_s(\mathbf{E+\mathbf{u_s}\times\mathbf{B}})																								\\
& \frac{\partial p_{s\parallel}}{\partial t}  + \nabla \cdot ( p_{s\parallel} \mathbf{u_s}) =    - 2 p_{s\parallel} \mathbf{b} \cdot (\mathbf{b} \cdot \nabla ) \mathbf{u_s} 	\\
& \frac{\partial p_{s\perp}}{\partial t} + \nabla\cdot( p_{s\perp}\mathbf{u_s}) = 
   - p_{s\perp} (\nabla \cdot \mathbf{u_s}) 
   + p_{s\perp} \mathbf{b} \cdot (\mathbf{b} \cdot \nabla ) \mathbf{u_s}
\end{align}
\end{subequations}
where $\rho$ and $\mathbf{u}$ denote the mass density and the velocity vector, respectively, and $q$ and $m$ are the charges and masses of the particles. 
 For convenience of implementation, we solve the average pressure $p=\frac{2 p_\perp + p_\parallel}{3}$ instead of the perpendicular pressure $p_\perp$ because $p$ is already a primitive variable solved by our MHD code BATS-R-US. The equation for $p$ can be obtained from combining equations \ref{eqn:sixmoment}c and \ref{eqn:sixmoment}d:
\begin{equation}
\label{eqn:sixmonent_p}
\frac{\partial p_s}{\partial t} + \nabla \cdot (p_s \mathbf{u}_s) = (p_s-p_{s\parallel}) \mathbf{b} \cdot (\mathbf{b} \cdot \nabla) \mathbf{u} 
    - \left(p_s-\frac{p_{s\parallel}}{3}\right) \nabla \cdot \mathbf{u}_s
\end{equation}

Alternatively, we can solve for the hydrodynamic energy density $e=\frac{\rho \mathbf{u} ^2}{2} + \frac{3}{2} p$ for each species:
\begin{equation}
\label{eqn:sixmonent_e}
\frac{\partial e_s}{\partial t} + \nabla \cdot [\mathbf{u_s} (e_s+p_s) + \mathbf{u_s} \cdot
(p_{s\parallel}-p_{s\perp}) \mathbf{b}\mathbf{b}] = \frac{q_s}{m_s}\rho_s \mathbf{u_s} \cdot \mathbf{E}
\end{equation}
which can be beneficial to get better jump conditions across shock waves. Note, however, that the parallel pressure equation is still solved with the adiabatic assumption, so non-adiabatic heating is not properly captured. In addition, the magnetic energy is not included into the energy density, so the jump conditions are only approximate. 
In general, there can be many more source terms on the right hand sides of the above equations 
corresponding to gravity, charge exchange, chemical reactions, collisions, etc.  

The electric field ($\mathbf{E}$) and magnetic field ($\mathbf{B}$) are obtained from the Maxwell equations:
\begin{subequations} 
\begin{eqnarray}
\label{eqn:maxwell_orig}
\frac{\partial \mathbf{B}}{\partial t} + \nabla \times \mathbf{E} &=& 0				\\
\frac{\partial \mathbf{E}}{\partial t} - c^2\nabla\times\mathbf{B} &=& -c^2\mu_0\mathbf{j} \\
\nabla \cdot \mathbf{E} &=& \frac{\rho_c}{\varepsilon_0}	\label{eqn:dive} \\
\nabla \cdot \mathbf{B} &=& 0 \label{eqn:divb}
\end{eqnarray}
\end{subequations}
where $\varepsilon_0$ is the vacuum permittivity, $\mu_0$ is the vacuum permeability, $c=1/\sqrt{\varepsilon_0 \mu_0}$ is the speed of light, $\rho_c = \sum_s (q_s/m_s)\rho_s$ is the total charge density and $\mathbf{j} = \sum_s (q_s/m_s) \rho_s \mathbf{u}_s$ is
the current density.

Equations~\ref{eqn:dive} and \ref{eqn:divb} are constraints on the initial conditions and analytically these conditions are preserved. Numerically, however, this is not guaranteed to hold. We use the hyperbolic/parabolic cleaning method \cite{Munz_1999, Munz_2000, Dedner_2002} to control the numerical errors in these equations. We introduce the scalars $\psi$ and $\phi$ as additional independent variables and solve the following modified form of the Maxwell equations:
\begin{subequations}
\label{eqn:maxwell}
\begin{eqnarray}
\frac{\partial \mathbf{B}}{\partial t} + \nabla \times \mathbf{E} + c_B \nabla \psi &=& 0							\\
\frac{\partial \mathbf{E}}{\partial t} - c^2 \nabla \times \mathbf{B} + c_E \nabla \phi &=&
     - c^2 \mu_0 \mathbf{j}	\\
\frac{\partial \psi}{\partial t} + c_B \nabla \cdot \mathbf{B} &=& -d_B \psi									\\
\frac{\partial \phi}{\partial t} + c_E \nabla \cdot \mathbf{E} &=& \frac{c_E}{\varepsilon_0} {\rho_c} - d_E \phi
\end{eqnarray}
\end{subequations}
where $c_B$ and $c_E$ are the hyperbolic propagation speeds, while $d_B$ and $d_E$ are the parabolic decay rates. 
To make the paper more self-contained, we provide a brief derivation in the Appendix to show how the hyperbolic/parabolic cleaning works for the six-moment equations.
On the other hand without using a cleaning method, Balsara et al. \cite{Balsara_2016} solved the magnetic and electric fields in plasma on a facially-collocated Yee-type mesh and proved that magnetic field is reconstructed in a divergence-free fashion and the electric field is reconstructed in a form that is consistent with the Gauss' law. 
Balsara et al. \cite{Balsara_2017, Balsara_2018} further extended their Yee-type mesh algorithm to simulate the electrodynamics in material media.

\section{Characteristic Wave Speeds}
\label{sec:wavespeed}

The fastest wave speed in the six-moment (also five- and ten-moment) equations is the speed of light $c$. Using $c$ in the numerical fluxes, however, makes the scheme rather diffusive. 
To reduce diffusion while maintaining stability, we use a point-implicit evaluation of the stiff source terms following Shumlak et al.\cite{Shumlak_2003}, who
proposed to ignore the Lorentz force terms and consequently
the interactions between the charged fluids and the electromagnetic fields while calculating the characteristic wave speeds.
Using this approach, the characteristic speed for each fluid will simply be its sound wave speed. 
We tried this approach, but found that it gives unsatisfactory results in several applications. 

We take an alternative approach by considering the wave speeds of MHD with anisotropic electron and ion pressures instead.
This takes into account fast magnetosonic waves, which is the proper wave speed in the MHD limit of the six-moment equations. 
On the other hand the electron sound speed and the whistler wave speed are not included, which reduces the numerical diffusivity,
and similarly to Shumlak et al.\cite{Shumlak_2003} we rely on the point-implicit scheme to provide numerical stability.
In the following discussion, we limit our derivation to a single ion fluid and a single electron fluid, 
and we use the subscript $i$ to denote the ion fluid while the subscript $e$ is for the electron fluid.  
The proper generalization to arbitrary number of fluids is left for future work 
(we currently employ some heuristic formulas that work reasonably in most cases but may not be valid in general).

It is important to note that the following equations in this section are only used to derive the characteristic speeds, but not used in the six-moment model at all. In the MHD approximation, the electric field is obtained from the electron momentum equation 
by ignoring the electron inertial terms, which gives
\begin{equation}
\label{eqn:efield}
\mathbf{E} = -\mathbf{u_e} \times \mathbf{B} - \frac{1}{e n_e} \nabla \cdot [p_{e\perp} \mathbf{I} + (p_{e\parallel} - p_{e\perp})\mathbf{bb}]
\end{equation}
The electron number density $n_e$ can be obtained from charge neutrality as $n_e = n_i q_i/e$ (or simply $n_i$ for singly charged ions). 
For the momentum equation, the electron velocity is expressed from the current density as $\mathbf{u_e} = \mathbf{u_i} - \mathbf{j}/(e n_e)$ resulting in 
the usual MHD Lorentz force $\mathbf{j}\times\mathbf{B}$ in the ion momentum equation.
The current density is obtained from Ampere's law (after dropping the displacement current)
as $\mathbf{j}=\nabla\times\mathbf{B}/\mu_0$ as usual in the MHD approximation. 
In all the other equations we take $\mathbf{u_e} = \mathbf{u_i}$, so the governing equations for the ions become
\begin{subequations}
\begin{align}
&\frac{\partial \rho}{\partial t} + \nabla \cdot (\rho \mathbf{u}) = 0 
\\
&\frac{\partial \rho \mathbf{u}}{\partial t}  + \nabla \cdot [\rho \mathbf{u}\mathbf{u} + p_{\perp} \mathbf{I} + (p_{\parallel} - p_{\perp} )\mathbf{bb}]
+ \frac{\mathbf{B}}{\mu_0} \times (\nabla \times \mathbf{B}) = 0
\\
&\frac{\partial p_{\parallel}}{\partial t}  + \nabla \cdot ( p_{\parallel}   \mathbf{u})                                                        + 2 p_{\parallel} \mathbf{b} \cdot (\mathbf{b} \cdot \nabla ) \mathbf{u} = 0 \\
&\frac{\partial p_{\perp}}{\partial t}       + \nabla \cdot ( p_{\perp}   \mathbf{u}) + p_{\perp} (\nabla \cdot \mathbf{u}) -    p_{\perp} \mathbf{b} \cdot (\mathbf{b} \cdot \nabla ) \mathbf{u}  = 0   
\end{align}
\label{eqn:sixmoment_mhd}
\end{subequations}
where $p_\perp = p_{i\perp} +  p_{e\perp} $, $p_\parallel = p_{i\parallel} + p_{e\parallel}$, $\rho = \rho_i$ and $\mathbf{u} = \mathbf{u_i}$.

The magnetic field can be obtained from the classical ideal MHD induction equation ignoring the Hall terms, which can be written as
\begin{equation}
\label{eqn:B_induction}
\frac{\partial \mathbf{B}}{\partial t} = \nabla \times (\mathbf{u} \times \mathbf{B})
\end{equation}

In a six-moment simulation, the speed of light is usually reduced to speed up the simulation, in which case the reduced 
speed of light need to be properly set to make sure that it must be larger than any of the characteristic speeds. 
In such a system, the characteristic speeds may not be much smaller than the reduced speed of light, so the semi-relativistic situation needs to 
be considered. 
In the semi-relativistic case, we only need to modify the momentum equation from the classical limit. The non-conservative form of the momentum equation
(Equation\,\ref{eqn:sixmoment_mhd}b) can be written as
\begin{equation}
\begin{split}
\rho  \frac{\partial \mathbf{u}}{\partial t}  & + \gamma_A^2 (\mathbf
{I} + \frac{V_A^2}{c^2} \mathbf{bb}) \cdot \{ \rho (\mathbf{u} \cdot \nabla) \mathbf{u} + \nabla p_{\perp} + \nabla \cdot [(p_{\parallel} - p_{\perp} )\mathbf{bb}] \} \\
& + \frac{\gamma_A^2}{\mu_0} \mathbf{B} \times [\nabla \times \mathbf{B} - \frac{1}{c^2} \mathbf{u} \times (\nabla \times \mathbf{E}) - \frac{1}{c_0^2}\mathbf{u} \nabla \cdot \mathbf{E}] = 0
\end{split}
\label{eqn:semirelativistic}
\end{equation}
where 
\begin{equation}
\gamma_A = \frac{1}{\sqrt{1+\frac{V_A^2}{c^2}}}
\end{equation}
is the Alfv\'en factor, $c_0$ is the true value of the speed of light and $c$ is the artificially reduced speed of light.
The term $c_0^{-2}\mathbf{u} \nabla \cdot \mathbf{E}$ can be dropped because $u$ is much smaller than $c_0$ for the semi-relativistic limit and
this term is much smaller than $\nabla \times \mathbf{B}$.

We want to obtain the characteristic wave speed of Equations (\ref{eqn:sixmoment_mhd}) and (\ref{eqn:B_induction}) with the ion momentum equation replaced by Equation \ref{eqn:semirelativistic}).  First we write the one dimensional (along the $x$ direction) equations  in the form $\frac{\partial \mathbf{U}}{\partial t} + \mathbf{M_x}\frac{\partial \mathbf{U}}{\partial x} = 0$ where $\mathbf{M_x}$ is the characteristic matrix.
In the MHD approximation the variable array reduces to $\mathbf{U} = (\rho, \mathbf{u}, \mathbf{B}, p_\parallel, p_\perp) = (\rho, u_x, u_y, u_z, B_x, B_y, B_z, p_\parallel, p_\perp)$.
In 1D $B_x$ is a constant , so the variable array can be further reduced to
$\mathbf{U} = (\rho, \mathbf{u}, B_y, B_z, p_\parallel, p_\perp)$. These variables only depend on $x$ and $t$.
We further simplify the problem by rotating the coordinate system such that the magnetic field is in the $x-y$ plane so that $B_z =0$. 
The characteristic matrix $\mathbf{M_x}$ is obtained by the Mathematica software:

$ \mathbf{M_x} = 
\left( 
\begin{array}{cccccccc}
 u_x & \rho                                            & 0                                                    & 0                                                    & 0                & 0                & 0                    & 0 \\
 0        & \gamma_A^2 u_x + \chi_{11}  & \chi_{12}                                       & \chi_{13}                                        & \kappa_1    & 0                & \eta_{12}       & \eta_{12} \\
 0        & \chi_{21}                                  & \gamma_A^2 {u_x} + \chi_{22}     & \chi_{23}                                        & \kappa_2   & 0                & \eta_{21}       & \eta_{22} \\
 0        & \chi_{31}                                   & \chi_{32}                                       & \gamma_A^2 {u_x}+\chi_{33}       & \kappa_3   & \nu             & 0                    & 0 \\
 0        & B_y                                           & -{B_x}                                            & 0                                                   & {u_x}          & 0               & 0                    & 0 \\
 0        & 0                                               & 0                                                    & -{B_x}                                           & 0                & {u_x}          & 0                    & 0 \\
 0        & p_\parallel(2 {b_x}^2 + 1)         & 2 {p_\parallel} {b_x} {b_y}             & 0                                                   & 0                 & 0               & {u_x}              & 0 \\
 0        & p_\perp (2 -{b_x}^2)                 & - {p_\perp} {b_x} {b_y}                   & 0                                                   & 0                 & 0               & 0                    & {u_x} \\
\end{array}
\right)
$ \\
where 

$\\$

$\chi = \frac{\gamma_A^2}{\mu_0 \rho c^2}
\left(
\begin{array}{ccc}
 (B_x^2 - B_y^2) u_x               & 2 B_x B_y u_x              & 0 \\
 2 B_x B_y u_x                        & (B_y^2 - B_x^2) u_x     & 0 \\
 -B_y^2 u_z                             & B_x B_y u_z               & -B_x^2 u_x - B_x B_y u_y \\
\end{array}
\right)
$

$\\$

$\kappa = \frac{\gamma_A^2}{\mu_0 \rho c^2}
\left(
\begin{array}{c}
  (c^2 - u_x^2) B_y + (2 \mu_0 \rho c^2 B^{-2} + 1) b_x^2 B_y (p_\perp - p_\parallel) \rho^{-1}                       \\
  (u_x^2 - c^2) B_x + (b_y^2 + (b_y^2 - b_x ^2)\mu_0 \rho c^2 B^{-2}) B_x (p_\perp - p_\parallel) \rho^{-1}  \\
 -B_y u_x u_z \\
\end{array}
\right)
$

$\\$

$\nu  = \frac{\gamma_A^2}{\mu_0 \rho c^2} [(u_x^2 - c^2) B_x + B_y u_x u_y - \mu_0 c^2 B_x B^{-2} (p_\perp - p_\parallel)]$

$\\$

$\eta = \frac{1}{\rho}
\left(
\begin{array}{cc}
   b_x^2      &   \gamma_A^2 b_y^2                  \\
   b_x b_y   &  - \gamma_A^2 b_x b_y                 \\
\end{array}
\right)
$

$\\$

The matrix $\mathbf{M_x}$ is identical to the submatrix (the upper left $8\times8$ elements) of the characteristic matrix 
that Meng et al. \citep{Meng_2012} (hereafter \textit{Paper I}) obtained with pressure anisotropy in ions and isotropy in electrons. We note that here the parallel and perpendicular pressures are the sums of the ion and electron pressures.
We also correct a typo in the second element of the matrix $\kappa$ in \textit{Paper I}, where $B_x^2$ should be $B_x$.
We use Mathematica to solve the characteristic equation $\det(\mathbf{M_x} - \lambda \mathbf{I}) =0$ and after some tedious algebra, the characteristic equation can be written as
\begin{equation}
 (\lambda - u_x)^2 \rm P_2(\lambda) P_4(\lambda) = 0
\end{equation}
where the wave speed $\lambda$ is one of the eigenvalues of $\mathbf{M_x}$ and $\rm P_2$ and $\rm P_4$ are second- and fourth-order polynomials, respectively: 
\begin{subequations}
\begin{align}
{\rm P_2} = & \lambda(\lambda - u_x) + \gamma_A^2 [\lambda (\mathbf{u} \cdot \mathbf{b}) \mathbf{b_x} \frac{V_A^2}{c^2} - u_x(\lambda - u_x) - (V_A^2 + \frac{p_\perp - p_\parallel}{\rho}) b_x^2]        \\
\begin{split}
{\rm P_4} = & (\lambda - u_x)^4 - (\frac{2 p_\perp}{\rho} + \frac{2 p_\parallel - p_\perp}{\rho}b_x^2)(\lambda - u_x)^2  - (c^2 - \lambda^2) \frac{V_A^2}{c^2}[(\lambda - u_x)^2 - \frac{3 p_\parallel}{\rho} b_x^2]\\
                   & - [\frac{p_\perp^2}{\rho^2}(1-b_x^2) - \frac{3 p_\parallel p_\perp}{\rho^2} (2-b_x^2) + \frac{3 p_\parallel^2}{\rho^2} b_x^2] b_x^2
\end{split}
\end{align}
\end{subequations}
where $V_A^2 = B^2/(\mu_0 \rho)$ is the square of the classical Alfv\'en speed.

The $\rm P_2$ and $\rm P_4$ polynomials are identical to the $\rm P_2$ and $\rm P_4$ expressions in \textit{Paper I} after substituing $p_e = 0$ for the isotropic electron pressure in \textit{Paper I}. There are, however, two typos in $\rm P_4$ in \textit{Paper I}. The correct expression should be (with the corrections highlighted in red):
\begin{equation}
\begin{split}
{\rm P_4} = & (\lambda - u_x)^4 - (a^2+\frac{2p_\perp - 3p_\parallel}{\rho}+\frac{2p_\parallel - p_\perp}{\rho}b_x^2)(\lambda - u_x)^2-(c^2-\lambda^2)\frac{V_A^2}{c^2}[(\lambda - u_x)^2-a^2b_x^2] \\
& -[\frac{p_\perp^2-3p_\perp p_\parallel}{\rho^2}(1-b_x^2)+\frac{3p_\parallel^2}{\rho^2}b_x^2 {\color{red} + }\frac{5p_e}{3\rho}(\frac{4p_\parallel-p_\perp}{\rho}b_x^2-\frac{3p_\parallel}{\rho}) \color{red}{ -\frac{3p_\parallel p_\perp}{\rho^2}}]b_x^2
\end{split}
\end{equation}
We note that these typos in \cite{Meng_2012} are only in the published paper, and the equations used in Maple to derive the wave speeds and the wave speeds implemented into the code are all correct.
There are eight eigenvalues for the characteristic equation and each one is associated with one characteristic wave. Two of the eigenvalues are straightforward:
\begin{equation}
\lambda_{1,2} = u_x
\end{equation}
which are the two entropy waves related to $p_\perp$ and $p_\parallel$. 

\subsubsection{Alfv\'en wave}

As the $\rm P_2$ polynomial is the same as in \textit{Paper I}, the roots corresponding to the Alfv\'en wave speeds are the same too:
\begin{equation}
\begin{split}
\lambda_{4,5} = & \frac{1}{2} \gamma_A^2 [u_x - \frac{V_A^2}{c^2} (\mathbf{u \cdot \mathbf{b}}) b_x] + \frac{u_x}{2}  \\
                          & \pm \sqrt{\gamma_{A}^2 \left(V_{A,x}^2 + \frac{p_\perp - p_\parallel}{\rho} b_x^2\right) 
+ \left[\frac{1}{2} \gamma_A^2 (u_x - \frac{V_A^2}{c^2} (\mathbf{u} \cdot \mathbf{b}) b_x) + \frac{u_x}{2}\right]^2}
\end{split}
\end{equation}
where $V_{A,x} = V_A b_x = \sqrt{B_x^2/(\mu_0 \rho)}$ is the classical Alfv\'en wave speed in the $x$ direction.
It is important to point out that even though this formula looks the same as the solution in \textit{Paper I}, the physical
meaning is not the same because in our case, $p_\perp$ and $p_\parallel$ are the sum of parallel and perpendicular
pressures of ions and electrons, which means that in our case, the electron pressure does contribute to the Alfv\'en wave speed.

In the classical limit ($V_A \ll c$ and $\gamma_A \rightarrow 1$), the solutions reduce to 
\begin{equation}
\lambda_{3,4} = u_x \pm \sqrt{\frac{B_x^2}{\mu_0 \rho} + \frac{p_\perp - p_\parallel}{\rho} b_x^2}
\label{eqn:alfven}
\end{equation}

\subsubsection{Fast and slow magnetosonic waves}

The exact solutions of $\rm P4$ are too complicated to obtain. We follow the approach suggested in \textit{Paper I} to obtain
the approximate fast and slow magnetosonic wave speeds. We first obtain the solutions in the classical limit, in which case $\rm P4$ simplifies to 
\begin{equation}
\begin{split}
{\rm P_4} = & (\lambda - u_x)^4 - \left(V_A^2 + \frac{2 p_\perp}{\rho} + \frac{2 p_\parallel - p_\perp}{\rho}b_x^2\right)(\lambda - u_x)^2  \\
                   & - \left[\frac{p_\perp^2}{\rho^2}(1-b_x^2) - \frac{3 p_\parallel p_\perp}{\rho^2} (2-b_x^2) + \frac{3 p_\parallel^2}{\rho^2} b_x^2 - \frac{3 p_\parallel}{\rho} V_A^2\right] b_x^2
\end{split}
\end{equation}

The solutions can be easily obtained as:
\begin{equation}
\begin{split}
\lambda_{5,6,7,8} = u_x & \pm \frac{1}{\sqrt{2 \rho}}\{ (\frac{B^2}{\mu_0} + {2 p_\perp} + (2 p_\parallel - p_\perp) b_x^2) \pm [ (\frac{B^2}{\mu_0} + {2 p_\perp} + (2 p_\parallel - p_\perp) b_x^2)^2 \\
& + 4({p_\perp^2} b_x^2 (1-b_x^2) - 3 p_\parallel p_\perp b_x^2 (2-b_x^2) + 3 p_\parallel^2 b_x^4 - 3 p_\parallel \frac{B_x^2}{\mu_0} ) ]^{1/2} \}^{1/2}
\end{split}
\end{equation}

The solutions look the same as the formula obtained in \textit{Paper I} (when neglecting the electron pressure) and Baranov et al. (1970) \cite{Baranov_1970}. We would like to 
correct another typo in $\lambda_{5,6,7,8}$ in \textit{Paper I}. The first term in the second line of the expression should be $p_\perp^2 b_x^2 (1-b_x^2)$. 
The complete correct expression is

\footnotesize
\begin{equation}
\begin{split}
\lambda_{5,6,7,8} = u_x & \pm \frac{1}{\sqrt{2 \rho}}\{ (\frac{B^2}{\mu_0} + {2 p_\perp} + \frac{5}{3}p_e + (2 p_\parallel - p_\perp) b_x^2) \pm [ (\frac{B^2}{\mu_0} + {2 p_\perp} +  \frac{5}{3}p_e + (2 p_\parallel - p_\perp) b_x^2)^2 \\
& + 4({p_\perp^2} b_x^2 (1-b_x^2) - 3 p_\parallel p_\perp b_x^2 (2-b_x^2) + 3 p_\parallel^2 b_x^4 + 
   \frac{5}{3}p_e(4p_\parallel b_x^2-p_\perp b_x^2 -3p_\parallel)b_x^2 - 3 (p_\parallel + \frac{5}{3}p_e) \frac{B_x^2}{\mu_0} ) ]^{1/2} \}^{1/2}
\end{split}
\end{equation}
\normalsize

The next step is to extend the above solutions to the semi-relativistic case by considering some special cases (for example, $\mathbf{u} = 0$ and $b_x = 1$). The steps are the same as in \textit{Paper I} (after setting the isotropic electron pressure to zero and adding the anisotropic electron pressure to the total pressure) so we do not repeat the procedure here. The final approximate formulas for the fast and slow wave speeds can be written as
\begin{subequations} \label{eqn:slowfast}
\begin{align}
\tilde{\lambda}_{5,6} = & u_x  \pm \tilde{c}_x = u_x \pm \frac{1}{\sqrt{2}} \sqrt{\gamma_A^2 (\overline{a}^2 + \overline{V}_A^2) - \sqrt{\gamma_A^4 (\overline{a}^2 + \bar{V}_A^2)^2 - 4 \gamma_A^2 (a^2 \overline{V}_{A,x}^2 + b^2)}} \\
\tilde{\lambda}_{7,8} = & \gamma_A^2 u_x  \pm \tilde{c}_f = \gamma_A^2 u_x \pm \frac{1}{\sqrt{2}} \sqrt{\gamma_A^2 (\overline{a}^2 + \overline{V}_A^2) + \sqrt{\gamma_A^4 (\overline{a}^2 + \bar{V}_A^2)^2 - 4 \gamma_A^2 (a^2 \overline{V}_{A,x}^2 + b^2)}}
\end{align}
\end{subequations}
where 
$\overline{a}^2 = a^2(1+\frac{V_{A,x}^2}{c^2}) + \frac{2 p_\perp - 3p_\parallel}{\rho} + \frac{2p_\parallel - p_\perp}{\rho}b_x^2$, 
$b^2                 = \frac{b_x^2}{\rho^2}[3 p_\parallel p_\perp (2-b_x^2) - p_\perp^2(1-b_x^2) - 3 p_\parallel^2 b_x^2] $, 
$ \overline{V}_A^2      = V_A^2(1-\gamma_A^2 \frac{u_x^2}{c^2}) $ and 
$\overline{V}_{A,x}^2 = V_{A,x}^2(1-\gamma_A^2 \frac{u_x^2}{c^2})$. \textit{Paper I} showed numerically that these approximate speeds are accurate in most of the practically important parameter regime.

\section{Discretization}
\label{sec:discretization}

In the following subsections, we describe how we discretize the fluxes (the pure divergence terms 
on the left hand side) and the source terms on the right hand side
of the six-moment equations.

The time step is limited by the Courant-Friedrichs-Lewy (CFL) condition based on the speed of light. In practice, we can reduce the speed of light to a value that is a factor of 2-3 faster than the fastest flow and fast wave speed obtained in the previous section to speed up the simulation.

\subsection{Source terms}

The stiff source terms are evaluated by a new point-implicit scheme. Only the momenta and the electric field are involved, so the implicit variables are
\begin{equation}
\mathbf{U}_{impl} = 
\left(
\begin{array}{c}
   \rho_s\mathbf{u_s}         \\
   \mathbf{E}                 \\
\end{array}
\right)
\end{equation}
In the momentum equations and the Maxwell equation for the electric field we split the various terms into two groups: the fluxes and non-stiff source terms $\mathbf{R}_{expl}$ and the stiff source terms $ \mathbf{S}_{impl}$ containing the Lorentz force terms in the momentum equations and the $c^2 \mu_0 \mathbf{j}$ term in the Maxwell equations. The stiff source terms can be written as:
\begin{equation}
\mathbf{S}_{impl}(\mathbf{U}_{impl}) = 
\left(
\begin{array}{c}
   \frac{q_s}{m_s}(\rho_s\mathbf{E} + \rho_s\mathbf{u_s} \times \mathbf{B})                  \\
   -c^2 \mu_0 \sum \frac{q_s}{m_s} \rho_s \mathbf{u_s}                 \\
\end{array}
\right)
\end{equation}
which shows that $\mathbf{S}_{impl}$ is linear in $\mathbf{U}_{impl}$. The implicit variables $\rho_s\mathbf{u_s}$ (an independent variable and could be denoted as $\mathbf{m_s}$) and $\mathbf{E}$ are multiplied with explicit variables $\rho_s$ and $\mathbf{B}$. The point-implicit update is respective to the time level $n$, which is shown in Equation\,(\ref{eqn:point_implicit}).
With this notation the six-moment equations for the $\mathbf{U}_{impl}$ variables can be written as 
\begin{equation}
\frac{\partial \mathbf{U_{impl}}}{\partial t} =\mathbf{R}_{expl} + \mathbf{S}_{impl} 
\end{equation}
We use the following steps to update the point implicit variables (the rest of the variables are updated with a simple explicit step): 
\begin{subequations}
\label{eqn:point_implicit}
\begin{align}
& \Delta \mathbf{U}^*_{impl} = \Delta t  \mathbf{R}_{expl}																										\\
& \Delta \mathbf{U}^{n+1}_{impl} = \Delta \mathbf{U}^*_{impl} + \Delta t \mathbf{S}_{impl}^n + \beta \Delta \frac{\partial \mathbf{S}_{impl}}{\partial \mathbf{U}_{impl}} \Delta \mathbf{U}^{n+1}_{impl}				\\
& \mathbf{U}^{n+1}_{impl} = \mathbf{U}^n_{impl} + \Delta \mathbf{U}^{n+1}_{impl}																							
\end{align}
\end{subequations}
where $0.5 \le \beta \le 1$ is the time centering parameter and $\mathbf{R}_{expl} = \mathbf{R}_{expl}(\mathbf{U}_{impl})$ is restricted to the point implicit variables $\mathbf{U}_{impl}$. The three steps can be combined into a single update:
\begin{equation}
\mathbf{U}^{n+1}_{impl} = \mathbf{U}^n_{impl} + \left(\frac{I}{\Delta t} - \beta \frac{\partial \mathbf{S}_{impl}}{\partial \mathbf{U}_{impl}}\right)^{-1} \left(\mathbf{R}_{expl}^n + \mathbf{S}_{impl}^n\right)
\end{equation}
where $I$ is the identity matrix and the matrix $({I}/{\Delta t} - \beta {\partial \mathbf{S}_{impl}}/{\partial \mathbf{U}_{impl}})$ is obtained analytically and inverted numerically. We note that $\mathbf{S}_{impl}$ is linear so it is very easy to calculate the partial derivatives. For example, for one ion and one electron fluids
\begin{equation}
\frac{\partial \mathbf{S}_{impl}}{\partial \mathbf{U}_{impl}} =
\left(
\begin{array}{ccccccccc}
   0   & \frac{q_i}{m_i}B_z   & -\frac{q_i}{m_i}B_y     & 0       & 0       & 0        & w_i  & 0   & 0             \\
 -\frac{q_i}{m_i}B_z   & 0    & \frac{q_i}{m_i}B_x      & 0       & 0       & 0        & 0   & w_i  & 0             \\
 \frac{q_i}{m_i}B_y & -\frac{q_i}{m_i}B_x     & 0       & 0       & 0       & 0        & 0           & 0     & w_i  \\
 0       & 0        & 0     & 0    & \frac{q_e}{m_e}B_z    & -\frac{q_e}{m_e}B_y       & w_e  & 0   & 0             \\
 0       & 0        & 0   & \frac{q_e}{m_e}B_z    & 0     & -\frac{q_e}{m_e}B_x        & 0   & w_e  & 0             \\
 0       & 0        & 0   & \frac{q_e}{m_e}B_y   & -\frac{q_e}{m_e}B_x     & 0         & 0          & 0      & w_e  \\
 r_i   & 0   & 0   & r_e   & 0   & 0   & 0   & 0   & 0                                      \\
 0  & r_i    & 0   & 0   & r_e   & 0   & 0   & 0   & 0                                      \\
 0   & 0   & r_i   & 0   & 0   & r_e   & 0   & 0   & 0
\end{array}
\right)
\end{equation}
where $r_i = -c^2\mu_0{q_i}/{m_i}$, $r_e = -c^2\mu_0{q_e}/{m_e}$, $w_i = \rho_i{q_i}/{m_i}$ and $w_e = \rho_e {q_e}/{m_e}$.

This particular discretization of the point-implicit scheme has a very important property: it preserves steady state independent of the time step. If $\mathbf{R}_{expl}^n + \mathbf{S}_{impl}^n=0$, then $\mathbf{U}^{n+1}_{impl}=\mathbf{U}^n_{impl}$ independent of the value of $\Delta t$. This property is crucial when the source terms are very stiff, as is the case here. Alternative forms of the point-implicit scheme that do not have this property can produce incorrect solutions. 

For a second-order in time scheme, we use the point-implicit update in both the predictor and corrector steps. In the predictor step the time step is $\Delta t/2$ and we set $\beta=1$, while in the corrector step we use the full time step $\Delta t$, the time centered value for $\mathbf{R}_{expl}^{n+1/2}$ and $\beta=1/2$ to get second order accuracy for the point-implicit term: 
\begin{subequations}
\begin{align}
&\mathbf{U}^{n+1/2}_{impl} = \mathbf{U}^n_{impl} + \left(\frac{I}{\Delta t/2} - \frac{\partial \mathbf{S}_{impl}}{\partial \mathbf{U}_{impl}}\right)^{-1} 
\left( \mathbf{R}_{expl}^n + \mathbf{S}_{impl}(\mathbf{U}^{n}_{impl}, \mathbf{U}^{n+\alpha_1}_{expl})\right) \\
&\mathbf{U}^{n+1}_{impl} = \mathbf{U}^n_{impl} + \left(\frac{I}{\Delta t} - \frac12 \frac{\partial \mathbf{S}_{impl}}{\partial \mathbf{U}_{impl}}\right)^{-1}
\left(\mathbf{R}_{expl}^{n+1/2} + \mathbf{S}_{impl}(\mathbf{U}^{n}_{impl}, \mathbf{U}^{n+\alpha_2}_{expl}) \right) 
\end{align}
\end{subequations}
where $\alpha_1 = 0$ or $1/2$ and $\alpha_2 = 1/2\ \text{or}\ 1$ depending on the time levels of the explicit variables being used. The $\alpha_2 = 1/2$ option will achieve second-order accuracy, while $\alpha_2 = 1$ is more robust but not perfectly second order accurate. Our current implementation uses $\alpha_1=1/2$ and $\alpha_2=1$, i.e. the already updated explicit variables. 

This two-stage scheme also has the steady state conserving property. 
In steady state, the explicit update does not change the explicit variables, so $\mathbf{U}_{expl}^{n+\alpha_1}=\mathbf{U}^n_{expl}$. For the implicit variables,
if $\mathbf{R}_{expl}^n + \mathbf{S}_{impl}(\mathbf{U}^{n}_{impl}, \mathbf{U}^{n+\alpha_1}_{expl}) = \mathbf{R}_{expl}^n + \mathbf{S}_{impl}^n =  0$, then $\mathbf{U}_{impl}^{n+1/2}=\mathbf{U}^n_{impl}$ in the first stage. In the second stage the explicit update does not change the explicit variables, so $\mathbf{U}_{expl}^{n+\alpha_2}=\mathbf{U}^n_{expl}$ and $\mathbf{R}_{expl}^{n+1/2} + \mathbf{S}_{impl}(\mathbf{U}^{n}_{impl}, \mathbf{U}^{n+\alpha_2}_{expl})= \mathbf{R}_{expl}^{n}+ \mathbf{S}_{impl}^n=0$, and consequently  $\mathbf{U}^{n+1}_{impl}=\mathbf{U}^n_{impl}$.

It is worth to mention that Balsara et al. \citep{Balsara_2016} applied multiple stages with their Runge-Kutta implicit-explicit (IMEX) methods and could achieve a more accurate implicit-source discretization than our method. Abgrall and Kumar \citep{Abgrall_2014} used a very similar implicit source treatment with $\frac{\partial \mathbf{S}_{impl}}{\partial \mathbf{U}_{impl}}$ taken from the time level $n+1$ and showed that their point-implicit treatment could unconditionally preserve positivity.

\subsection{Physical fluxes}

The left hand sides of Equations (\ref{eqn:sixmoment} - \ref{eqn:sixmonent_e}) and (\ref{eqn:maxwell}) contain pure divergence terms and are obtained by the local Lax-Friedrichs or Rusanov scheme \citep{Rusanov_1961}:
\begin{equation}
\label{eqn:lax_scheme}
\frac{U_i^{n+1} - U_i^n}{\Delta t} = - \frac{F_{i+1/2}^n - F_{i-1/2}^n}{\Delta x} + \frac{a_{i+1/2}(U_{R,i+1/2}^n - U_{L,i+1/2}^n) - a_{i-1/2}(U_{R,i-1/2}^n - U_{L,i-1/2}^n)}{2 \Delta x} 
\end{equation}
where $U$ is one of the conservative variables, $F_{i+1/2}= \frac{F_{i+1/2}(U_R)+F_{i+1/2}(U_L)}{2}$ is the corresponding flux and $a$ is the fastest wave speed. The superscripts $n$ and $n+1$ refer to the time levels, while the subscript $i$ correspond to the spatial grid indexes. The subscript $i+1/2$ is the cell face between the cell centers $i$ and $i+1$, while the $R$ and $L$ subscripts correspond to the right and left extrapolated face values using some standard TVD type slope limiter. The second term on the right hand side is the numerical diffusion providing stability. In the standard Rusanov scheme 
the $a$ coefficients are set to the fastest wave speed (for the local state variables), which in principle should be the speed of light $c$. However, the use of speed of light to numerically diffuse all the variables greatly increases the numerical diffusion, and it requires a very fine grid and high computational cost to obtain an accurate solution.

As an alternate approach, we find that it is sufficient to use the fast magnetosonic speeds from Equation \ref{eqn:slowfast} and set $a_{i+1/2} = \max(|\lambda_7|, |\lambda_8|)$ for the plasma quantities $ \rho_s$, $\mathbf{u}_s$, $p_s$ and $p_{\parallel,s}$, the magnetic field $\mathbf{B}$ and its hyperbolic cleaning variable $\psi$,  while the electric field $\mathbf{E}$ and its hyperbolic cleaning
variable $\phi$ need to be numerically diffused by the speed of light using $a_{i+1/2}=c$.
These variable dependent wave speeds can substantially
reduce the numerical diffusion for the plasma quantities and the magnetic field and improve the accuracy of the solution
substantially.
However, the numerical diffusion algorithm proposed here may not work for super thermal electrons, 
while their sound speeds can be very large compared to cold or warm electrons. In such
a case, we provide an option to numerically diffuse a sub set of the variables with the speed of light, for example the electron density, momentum and pressure may be diffused with the speed of light, if necessary.

\subsection{Relaxation towards isotropy}
\textit{Paper I} discussed three kinds of instabilities (fire hose, mirror and proton cyclotron instabilities), which will
push the pressure tensor towards isotropy in the context of a single anisotropic ion fluid. As our model contain multiple fluids, the stability criteria become much more complicated. For sake of simplicity, we implemented a simple exponential decay term for each fluid (a right-hand-side source term in Equation (\ref{eqn:sixmoment}c), similar to that suggested in \textit{Paper I}:
\begin{equation}
\frac{\delta p_{s \parallel}}{\delta t} = \frac{p_s-p_{s \parallel}}{\tau_s}
\end{equation}
where $\tau_s$ is the relaxation time which relaxes the $p_\parallel$ towards $p$. In the extreme case when $\tau$ is extremely small
the anisotropy will relax to isotropy immediately, in which case the six-moment simulation becomes a five-moment simulation.

The source term is applied in the same way numerically as discussed in \textit{Paper I}, in a 
split manner at the end of the time step and discretized point-implicitly for the sake of numerical stability:
\begin{equation}
p_{s \parallel}^{n+1} = p_{s \parallel}^* + \frac{(p_s-p_{s \parallel}^*)\Delta t}{\Delta t + \tau}
\end{equation}
where $\Delta t$ is the stable time step, $*$ and $n+1$ are the incomplete and final time levels.

\section{Numerical Tests}
\label{sec:tests}

We perform a number of numerical tests to verify the robustness of the six-moment solver, including the light wave test to check the magnetic and electric field solver, the
fast wave test to check the propagation of the fast magnetosonic wave.
We also perform the GEM reconnection challenge \cite{Birn_2001} to test the 
applicability of the six-moment model to simulate the magnetic reconnection.
All the tests are performed in normalized units.

\subsection{Light Wave}
\begin{figure}[h!]
\center
\includegraphics[width=1.0\linewidth]{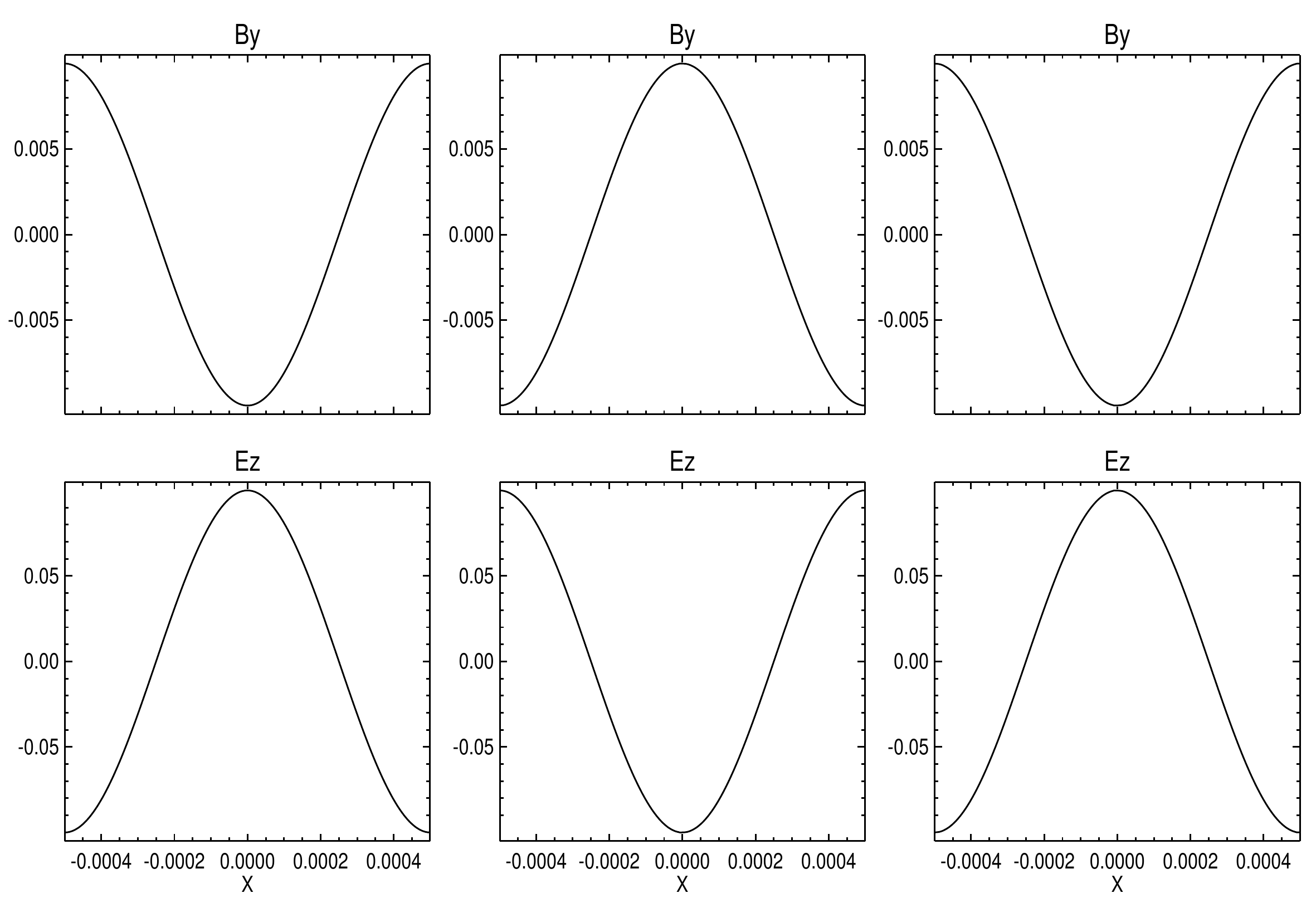}
\caption{Light wave propagation test. The upper panels show the evolution of $B_y$ at $t=0, 5\times10^{-5}\ \text{and } 1\times10^{-4}$, respectively; while the lower panels plot the $E_z$ component. The size of the domain is $10^{-3}$ so it takes $10^{-4}$ for a light wave with its speed of 10 to complete a period.}
\label{fig:light}
\end{figure}

The light wave test is set up on a 1-D grid between $x = \pm 5 \times 10^{-4}$ with periodic boundary conditions. The ion and electron masses are $m_i = 1$ and $m_e = 0.01$, respectively.
The initially uniform fluid states are set to $\rho_i = 1$, $\mathbf{u_i} = 0$, $p_{i, \perp} = p_{i, \parallel} = 5 \times 10^{-6}$, $\rho_e = 0.01$, 
$\mathbf{u_e} = 0$, $p_{e, \perp} = p_{e, \parallel} = 5 \times 10^{-6}$. We set the speed of light to $c=10$ and 
the magnetic and electric fields are perturbed with sinusoidal waves as $\delta B_y = - 0.01 \cos(2000\pi x)$ 
and $\delta E_z = 0.01 \cos(2000\pi x)$.
Figure \ref{fig:light} shows the simulation results with 800 grid cells after the wave propagating one full period showing the expected solution.

\begin{figure}[h!]
\center
\includegraphics[width=0.7\linewidth]{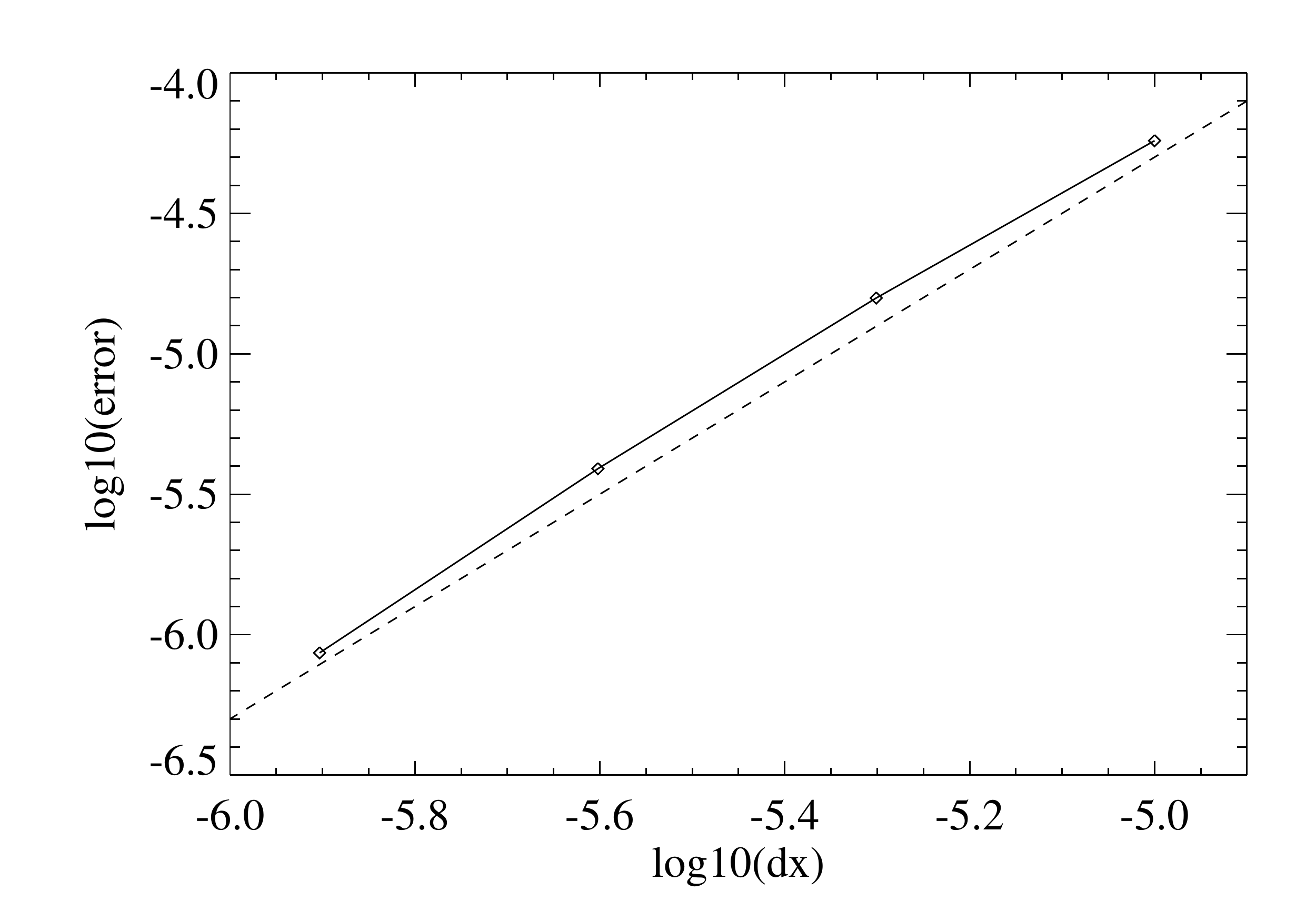}
\caption{The diamond-solid line shows the convergence rate for the light wave test while the dashed line shows the 2nd order convergence rate.}
\label{fig:conv_light}
\end{figure}

We did a grid convergence study with
$n_x=$ 100, 200, 400 and 800 grid cells using the 2nd order Rusanov scheme. The errors are calculated as the $L_1$ norm of the difference of the solution after 1 period relative to the initial condition.  
Figure \ref{fig:conv_light} shows the 
grid convergence rate for $B_y$, which is very close to the 2nd order convergence rate, as expected.

\subsection{\textit{Brio-Wu} Shock}

We carry out the \textit{Brio-Wu} shock \citep{Brio_1988} test with isotropic pressure for both ions and electrons (five-moment) 
in this session. The five-moment equations are chosen so the solution can be compared with published results \citep{Hakim_2006}, and also because the exact solution of the Brio-Wu shock for the six-moment system is unknown and depends on the pressure anisotropy behind the shock, which is not determined by conservation laws. At this time we have no physics based relaxation of the pressure anisotropy implemented for the six-moment equations.

\begin{table}[h!]
\centering
\begin{tabular}{ c | c  c}
\hline
                    & Left & Right     \\ \hline
 $\rho_i$      & 1      & 0.125     \\
 $\mathbf{u_i}$         & 0       & 0           \\
 $p_i$          & $5\times 10^{-5}$       & $5\times 10^{-6}$           \\ \hline
 $\rho_e$     & 1 $\cdot \frac{me}{mi}$      & 0.125$\cdot \frac{me}{mi}$     \\
 $\mathbf{u_e}$         & 0       & 0           \\
 $p_e$          & $5\times 10^{-5}$       & $5\times 10^{-6}$           \\ \hline
 $B_x$         & $0.75\times 10^{-2}$       & $0.75\times 10^{-2}$           \\
 $B_y$         & $1\times 10^{-2}$           & $-1\times 10^{-2}$           \\
 $B_z$         & 0       & 0           \\ \hline
 $\mathbf{E}$         & 0       & 0           \\
\hline
\end{tabular}
\caption{Initial conditions for the \textit{Brio-Wu} Shock test.}
\label{tab:shock}
\end{table}

\begin{figure}[h!]
\center
\includegraphics[width=1.0\linewidth]{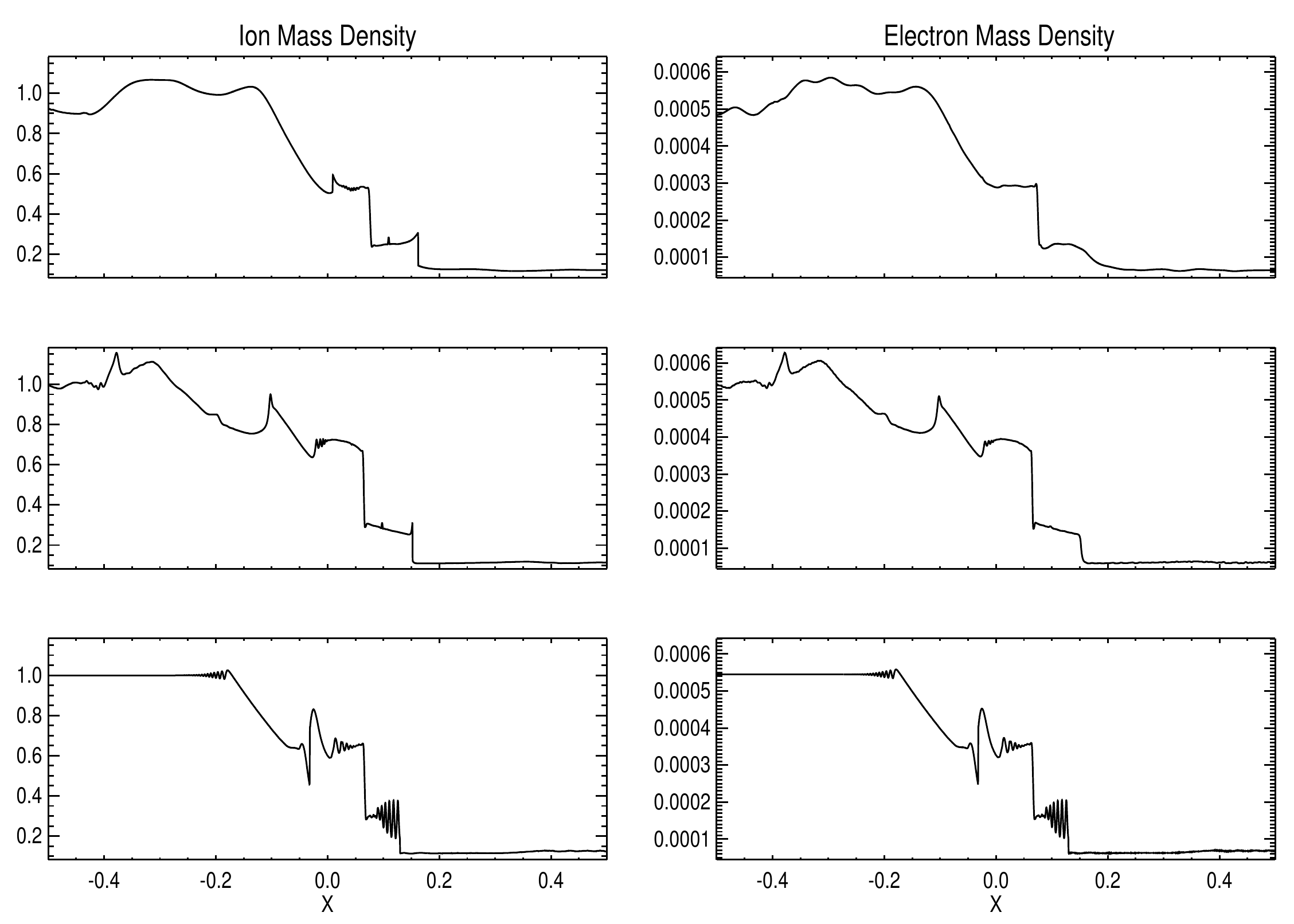}
\caption{Ion and electron mass densities at $t=10$ for the \textit{Brio-Wu} shock test. 
The upper panels are for $m_i = 1$, the middle panels are for $m_i = 0.1$, 
while the lower panels are for $m_i = 0.001$, respectively.}
\label{fig:shock}
\end{figure}

The test is set up on a 1-D grid between $x = \pm 0.5$ with $10^4$ cells and
open boundary conditions.
As suggested by Hakim et al. \cite{Hakim_2006}, 
the ion inertial length plays an important role when the electron fluid
is taken into account. 
So we set the ion mass $m_i$ to 1, 0.1, and 0.001, respectively, while
the ion mass to electron mass ratio is fixed (${m_i}/{m_e} =1836$).
The initial conditions are listed in Table\,\ref{tab:shock}.
Figure\,\ref{fig:shock} shows the simulation results at $t=10$. As expected, these results
are very close to the results published in Hakim et al. \cite{Hakim_2006}, except that there is a spike in the ion density at about $x=0.1$ for $m_i = 1$ and $m_i = 0.1$, and the oscillations do not start next to the shock at about $x=0.05$ for $m_i = 0.001$. The small differences may come from the different schemes and/or the point-implicit evaluations.

\subsection{Firehose instability}
We perform a test of the firehose instability on a 1-D grid between $x = \pm 6$ with $10^4$ cells and periodic boundary conditions.
We apply similar parameters as suggested in \textit{Paper I}, which is
$\rho_i = 1$, $\mathbf{u_i} = \mathbf{u_e} = 0$, $B_x = 10$, $B_y = B_z =0$, $\mathbf{E} = 0$, 
$p_{i, \parallel} = p_{e, \parallel} = 52$, $p_{i, \perp}  = p_{e, \perp} = 55/3$.
Due to the relatively small characteristic length, we set
$m_i = 0.001$, in which case the ion inertial length is much smaller than the characteristic length
so that the result is close to the classical MHD limit. The ion mass to electron mass ratio ${m_i}/{m_e}$
is set to 1000 and $n_e = n_i$ due to quasi-neutrality. 
We impose small perturbations on the background as 
$\delta u_{i,y} = \delta u_{e,y} = 0.01 \cos(k_A x) = 0.01 \cos(2 \pi x /6)$, 
$\delta B_y = 0.1 \cos(2 \pi x /6 + \pi/2)$ and the electric field is perturbed with the relation
$\mathbf{E} = - \mathbf{u_e} \times \mathbf{B}$.
Because the Alfv\'en speed ($v_A$) is $\sqrt{-1}$, which is obtained from Equation\,(\ref{eqn:alfven}), 
so the perturbations will not propagate but start to grow exponentially with $\exp(|{v_A}| k_A t)$.

\begin{figure}[h!]
\center
\includegraphics[width=0.7\linewidth]{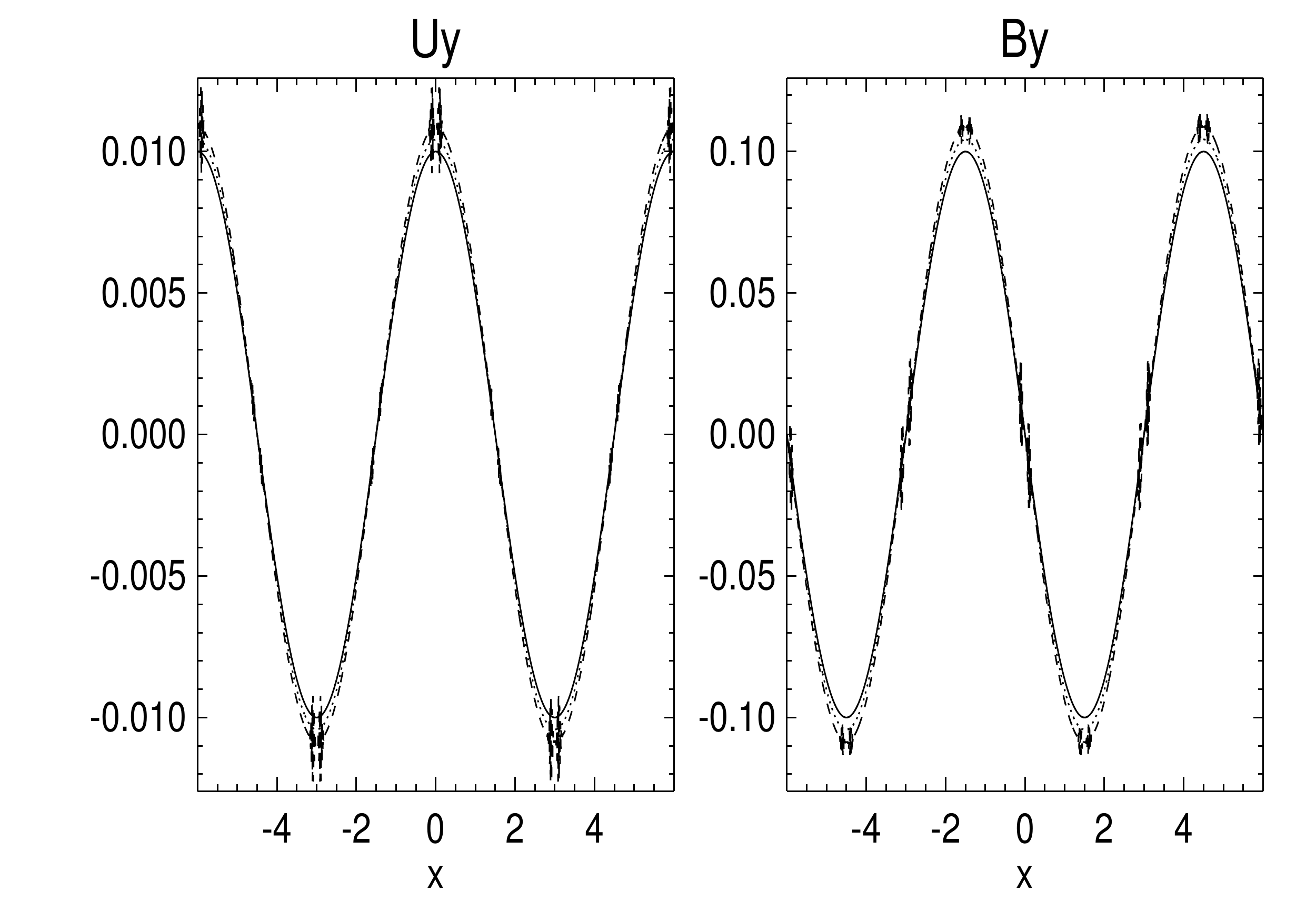}
\caption{$u_{i,y}$ and $B_y$ at different times. The solid line is at $t=0$, the dotted line
is at $t=0.04$ and the dashed line is at $t=0.08$. }
\label{fig:firehose}
\end{figure}

\begin{figure}[h!]
\center
\includegraphics[width=0.7\linewidth]{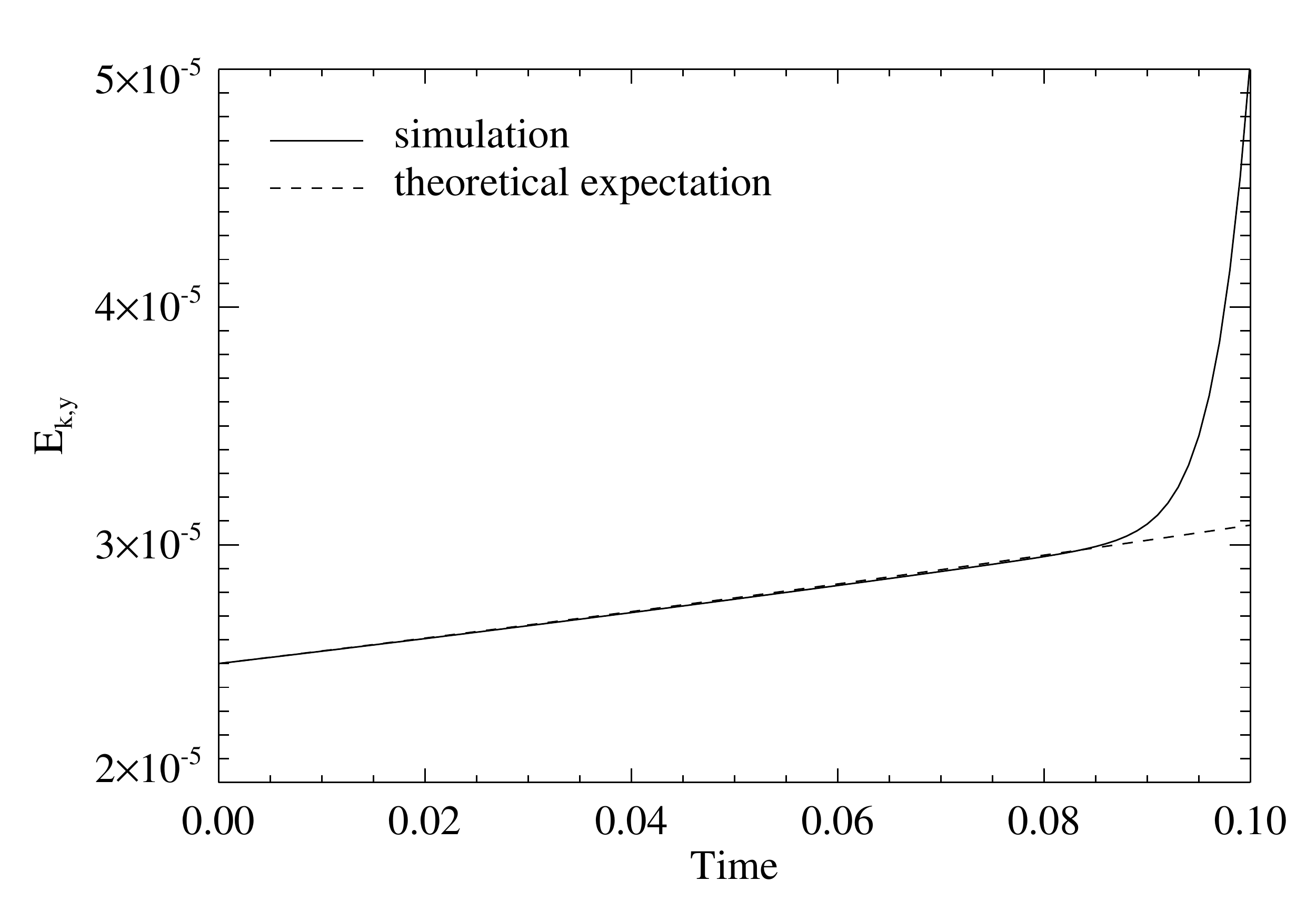}
\caption{The solid line plots the simulated kinetic energy in the $y$ direction $E_{k,y}$ while the dashed line
shows the theoretical $E_{k,y}$.}
\label{fig:firehose_growth_rate}
\end{figure}

Figure\,\ref{fig:firehose} plots $u_{i,y}$ and $B_y$ at three different times, which shows the growth
patten of the firehose instability. There are fast growing oscillations near the local extrema and sign change, which are
short wave length perturbations caused by the numerical errors. In the six-moment model,
ion and electron kinetics as well as point-implicit source terms are involved, which make the
six-moment model more complicated than the ideal anisotropic MHD. So it is not unexpected that
the short wave length perturbations appear much sooner than in the ideal anisotropic MHD test presented in \textit{Paper I}.
Figure\,\ref{fig:firehose_growth_rate} shows the agreement of the simulated growth rate of the average kinetic energy
$E_{k,y} = \rho_i u_{i,y}^2 /2$
and the analytical growth rate ($2 |{v_A}| k_A $, as the 
perturbations grow exponentially with $\exp(|{v_A}| k_A t)$)
is good until $t=0.085$. After $t=0.085$ the short wavelength perturbation becomes significant
and the growth rate of the kinetic energy deviates from the theoretical expectation.

\subsection{Fast Magnetosonic Wave}
Daldorff et al. \cite{Daldorff_2014} used the initial conditions 
\begin{subequations}
\label{eqn:fastwave}
\begin{align}
& n_i = n_0[1+ \delta \sin (k x- \omega t)] 						\\
& u_{i,x} =  c_f \delta \sin (k x- \omega t)						\\
& u_{i,y} = u_{i,z} 0										\\
& p_i = p_0 [1 + \gamma \delta \sin (k x- \omega t)]				\\
& p_{i,\parallel} = p_0 [1 +  \delta \sin (k x- \omega t)]				\\
& B_x = B_z = 0										\\
& B_y = B_0 [1+ p_0  \delta \sin (k x- \omega t)]
\end{align}
\end{subequations}
to set up a fast magnetosonic wave for the MHD equations with anisotropic ion pressure . 
Here $c_f = \frac{\omega}{k} = \sqrt{\frac{B_0^2 + 2 p_0}{\rho_{i, 0}}}$ is the fast wave propagation speed
moving perpendicular relative to the magnetic field direction and $\gamma = \frac{5}{3}$ is the adiabatic index. 
This is an exact solution for infinitesimal perturbation $\delta$.

\begin{figure}[h!]
\center
\includegraphics[width=1.0\linewidth]{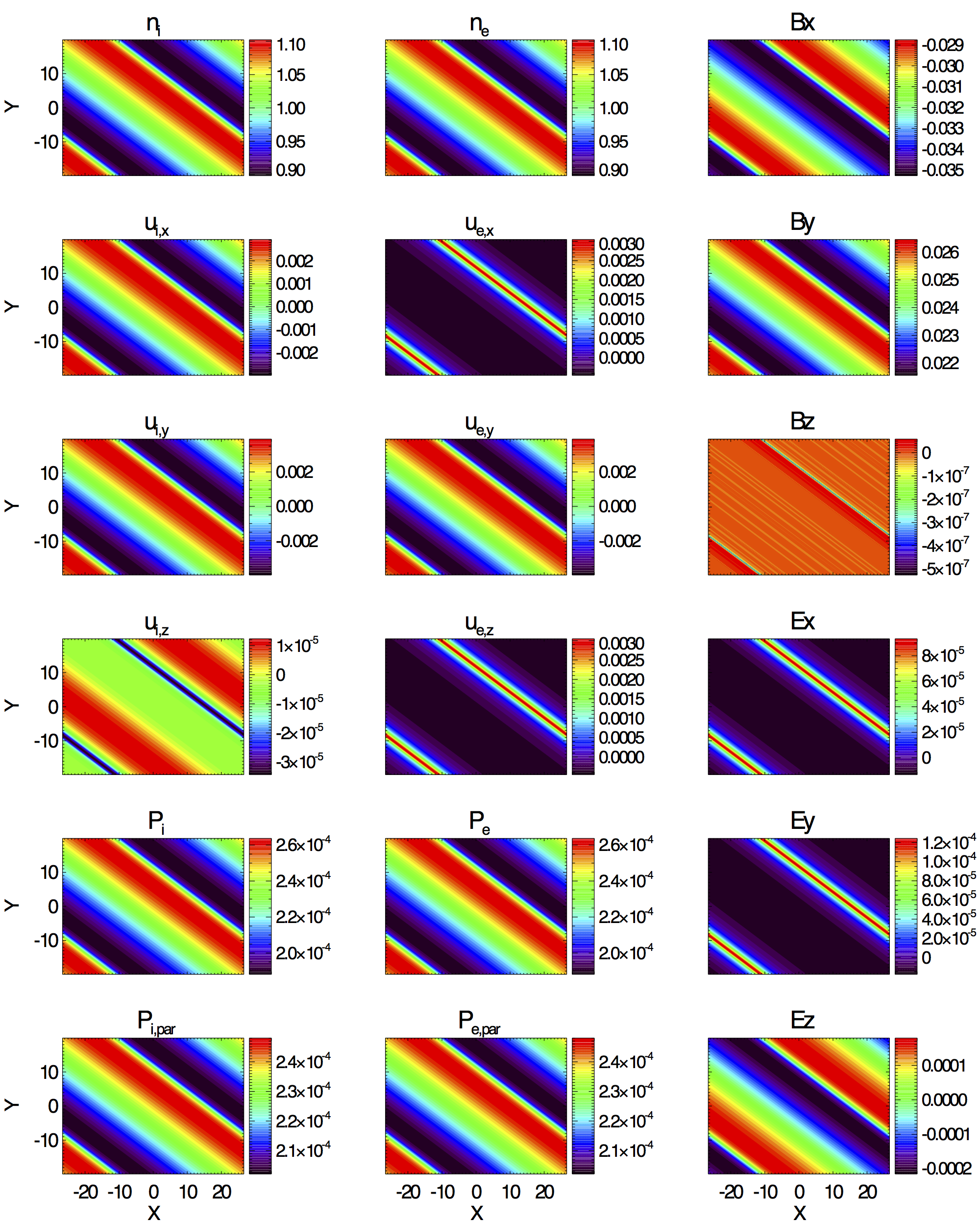}
\caption{Fast magnetosonic wave test on a 2-D grid at $t=512$. The left column shows the ion number density, velocity components, the scalar and parallel pressures. The middle
column shows the same variables for the electrons. The right column displays the magnetic and electric field components.}
\label{fig:fast_2d}
\end{figure}

\begin{figure}[h!]
\center
\includegraphics[width=1.0\linewidth]{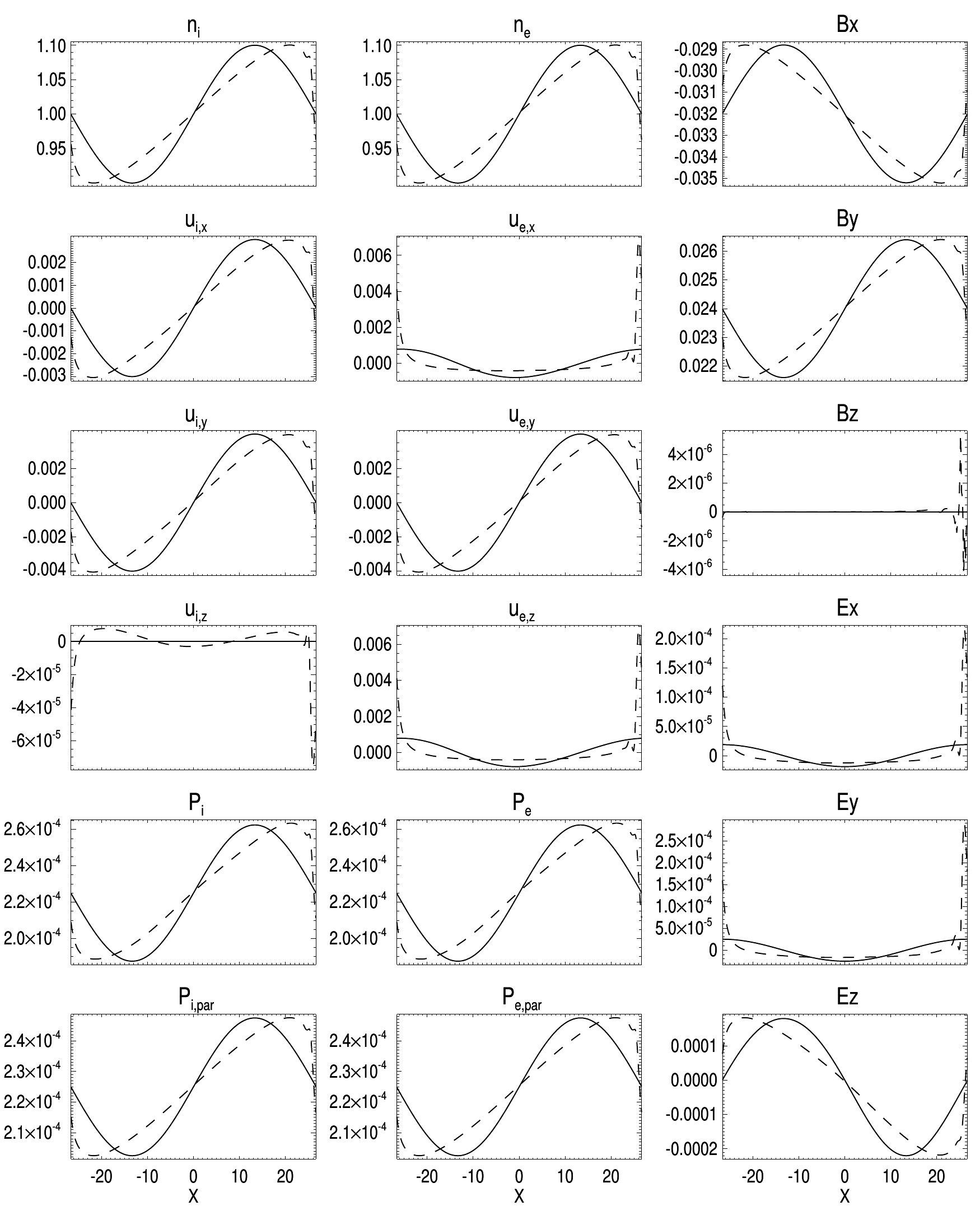}
\caption{Fast magnetosonic wave test on a 2-D grid at $t=0$ (solid line) and $t=640$ (dashed line) after the fast magnetosonic wave just finished one period.
Each panel shows the same variables as Figure \ref{fig:fast_2d} along the $x$ axis. The steepening of the fast
magnetosonic wave is well captured and it is similar as described by Daldorff et al. \citep{Daldorff_2014}.}
\label{fig:fast_1d}
\end{figure}

We follow Daldorff et al. \citep{Daldorff_2014} to create a test for the six-moment model.
When deriving the characteristic wave speeds in the classical MHD limit, we showed that the ion and electron pressures can be combined when pressure anisotropy exists in both ions and electrons. This means that we can adopt the above initial conditions by simply splitting the pressure evenly between ions and electrons.
\begin{subequations}
\label{eqn:fastwave}
\begin{align}
& p_i = p_e = \frac{p_0}{2} [1 + \gamma \delta \sin (k x- \omega t)]				\\
& p_{i,\parallel} = p_{e,\parallel} = \frac{p_0}{2} [1 +  \delta \sin (k x- \omega t)]
\end{align}
\end{subequations}
The simulation domain is on a 2-D grid bounded between $-80/3 < x < 80/3$ and $-20 < y < 20$. A single full wave with a rotation $\phi = \tan^{-1}(4/3)$ relative to the $x$ axis is used, which means that the wavelength is $\lambda = 32$, $k = 2 \pi / \lambda \approx 0.1964$ and $T\ \text{(period)}= \lambda/c_f = 640$ .
We set $n_0 = 1$, $m_i = 1$, $m_e = 0.01$, $p_0 = 4.5 \times 10^{-4}$, $B_0 = 0.04$, which gives the fast magnetosonic speed as $c_f = 0.05$.
The perturbation amplitude $\delta$ is set to 0.1, which is moderately non-linear, and the speed of light is $c=1$ (much larger than $c_f$ so that the system is in the classical limit). 
Figure \ref{fig:fast_2d} and \ref{fig:fast_1d} show the propagation of the simulated fast magnetosonic wave
on the 2-D grid with 512 cells along the $x$ direction and 384 cells along the $y$ direction. 
The simulation results are consistent with Daldorff et al. \citep{Daldorff_2014}.

\begin{figure}[h!]
\center
\includegraphics[width=0.7\linewidth]{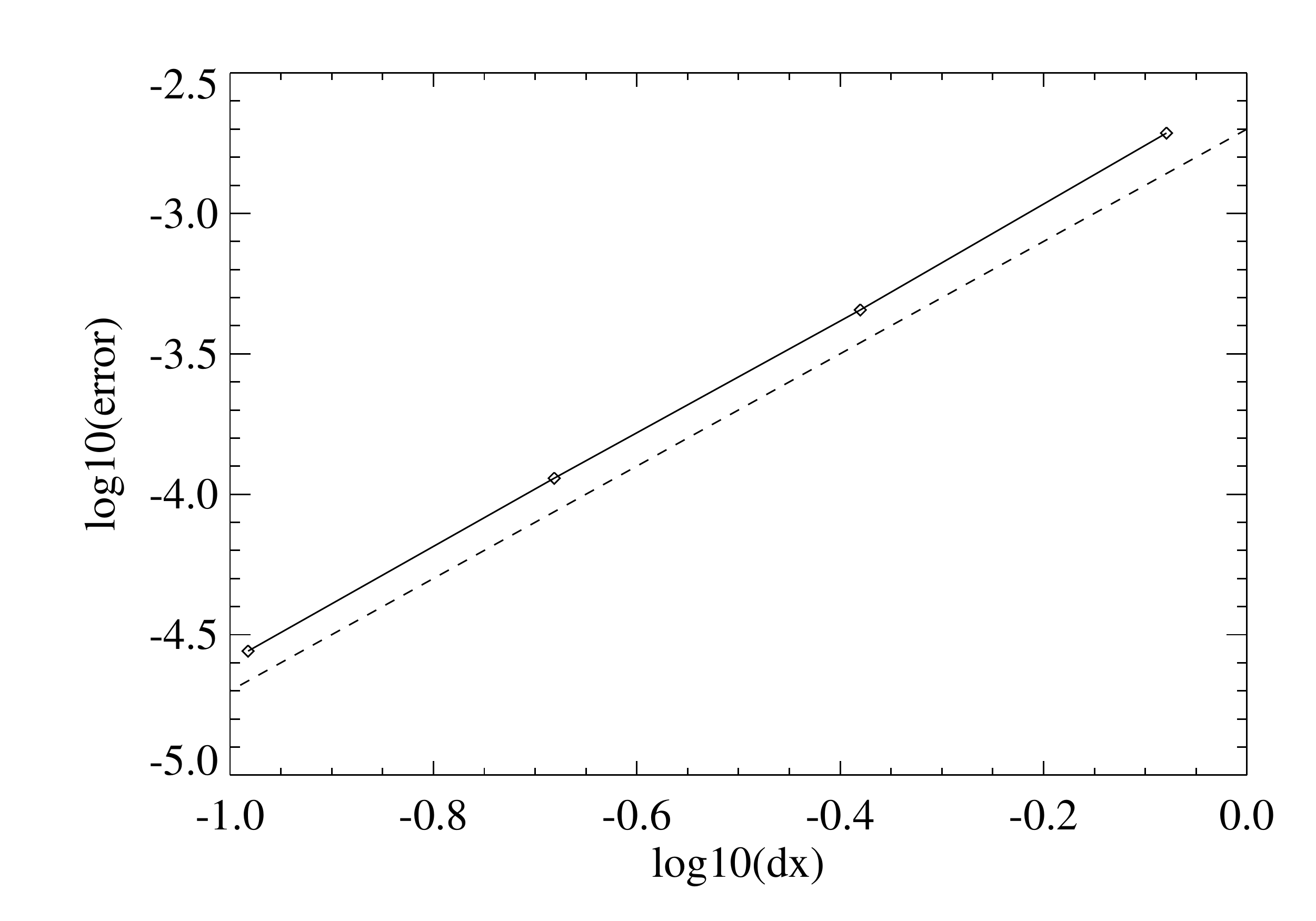}
\caption{The diamond-solid line shows the convergence rate for the fast wave test while the dashed line shows the 2nd order convergence rate.}
\label{fig:conv_fast}
\end{figure}

We carry out a grid convergence study for the fast wave test on the 2-D grid with 
$n_x=$ 64, 128, 256, 512 and 1024 cells along the $x$ direction and $n_y=$48, 96, 192, 384 and 768 cells
along the $y$ direction, respectively, using the 2nd order Rusanov scheme.
As we don't have an analytical solution for this non-linear fast wave, we use the simulation on the grid with 
$n_x=1024$ and $n_y=768$ as a reference solution and calculate the relative error of $E_z$. 
Figure\,\ref{fig:conv_fast} shows the grid convergence rate, which is very close to the 2nd order, as expected. We note that this test exercises the complete set of equations and their discretization, including the point-implicit scheme.

\subsection{GEM Reconnection}
The Geospace Environmental Modeling (GEM) reconnection challenge \cite{Birn_2001} has been widely applied to test a physics model's capability to simulate
the reconnection process. Many models, including resistive MHD \cite{Birn_2001}, Hall MHD \cite{Toth_2008, Birn_2001, Ma_2001}, hybrid models \cite{Birn_2001} and
Particl-In-Cell (PIC) models \cite{Hesse_2001, Pritchett_2001}, have been compared. 
The general conclusion of  \cite{Birn_2001} was that including the Hall physics is the minimum requirement to
correctly capture the fast magnetic reconnection rate.

The five-, six- and ten-moment multi-fluid models all include the Hall effect automatically by solving the exact Maxwell equations and allowing for different electron and ion velocities.
Hakim et al. \cite{Hakim_2006} carried out the GEM reconnection challenge and obtained some complex flow features in the electron fluid. 
Wang et al. \cite{Wang_2015} compared their five-moment and ten-moment two-fluid plasma model with a Particle-In-Cell (PIC)
model and showed that their five- and ten-moment models can reasonably reproduce some of the important electron kinetic features
observed in the PIC simulation during magnetic reconnections.
In this section, we perform six-moment simulations for the GEM reconnection challenge. 

The classical GEM reconnection challenge is based on the Harris current sheet equilibrium model, and such equilibrium only occurs when the inertial terms from the electrons are neglected. The classical 
Harris current sheet is not in equilibrium state for multi-fluid with electron fluid and particle-in-cell models. 
Eventhough we can obtain similar results as
Hakim et al. with our five-moment model, their initial conditions are obtained from the classical MHD limit and 
it is not applicable to the multi-fluid model because their initial conditions without perturbations are not in equilibrium state. 

We obtain the initial conditions starting from the oppositely directed magnetic fields
\begin{equation}
B_x = B_0 \tanh{\frac{y}{\lambda}} 
\end{equation}
where $B_0$ is the background magnetic field and $\lambda$ is the width of current sheet. 
The current density is then given by
\begin{equation}
J_z = -\frac{\partial B_x}{\partial y} = - \frac{B_0}{\lambda} \mathrm{sech}^2{\frac{y}{\lambda}} 
\end{equation}

Multiple equilibrium states can exist with different plasma conditions. We choose a uniform ion fluid background ($n_i$ and $p_i$ are uniform) with $u_i = 0$. There is no charge separation initially, so $n_e = n_i$ and $\mathbf{E} =0$. The ion fluid is in equilibrium in the unperturbed system. The current is carried by the electrons with the velocity
\begin{equation}
u_{e,z} = - \frac{J_z}{n_e e} = \frac{1}{n_e e} \frac{B_0}{\lambda} \mathrm{sech}^2{\frac{y}{\lambda}}   
\end{equation}
Substituting $n_e$, $\mathbf{u_e}$, $\mathbf{E}$ and $\mathbf{B}$ into the electron momentum equation, the electron fluid is in equilibrium when $p_{e, \parallel} = p_{e, \perp} = p_{e,0} + 0.5\cdot(B_0^2-B_x^2)$, where $p_{e,0}$ is the background electron pressure.

We set the background plasma parameters as 
$B_0 = 0.07$, $\lambda = 0.5$, $m_i = 1$, $m_e = 1/25$, $n_i = n_e = 1.225$, $\mathbf{u_i} = 0$
$p_{i, \perp} = p_{i, \parallel} =2.45 \times 10^{-3}$ and $p_{e,0} = 4.9 \times 10^{-3}$. The speed of light is set to $c=1$.
The simulation domain is a 2-D grid between $\pm L_x =  \pm 25.6$ in the $x$ direction with 512 cells and $\pm L_y = \pm12.8$ in the $y$ direction with 256 cells 
giving the grid resolution $\Delta x = \Delta y = 0.05$. Both the ion inertial length $d_i = \frac{1}{e}\sqrt{\frac{m_i}{\mu_0 n_i}} \approx 0.90$
and the electron skin depth $d_e = \frac{1}{e}\sqrt{\frac{m_e}{\mu_0 n_e}} \approx 0.18$ are reasonably well resolved. A periodic boundary
condition is applied in  the $x$ direction, while a reflecting boundary condition is applied in the $y$ direction.
We apply the same form of perturbation to the magnetic field $\delta \mathbf{B} = \mathbf{e_z} \times \nabla \chi$ as suggested by 
Birn et al. \cite{Birn_2001}, which is $\chi = \chi_0 \cos(2\pi x/ L_x)\cos(\pi x /L_y)$ and $\chi_0 = 0.1 B_0$.

\begin{figure}[h!]
\center
\includegraphics[width=0.65\linewidth]{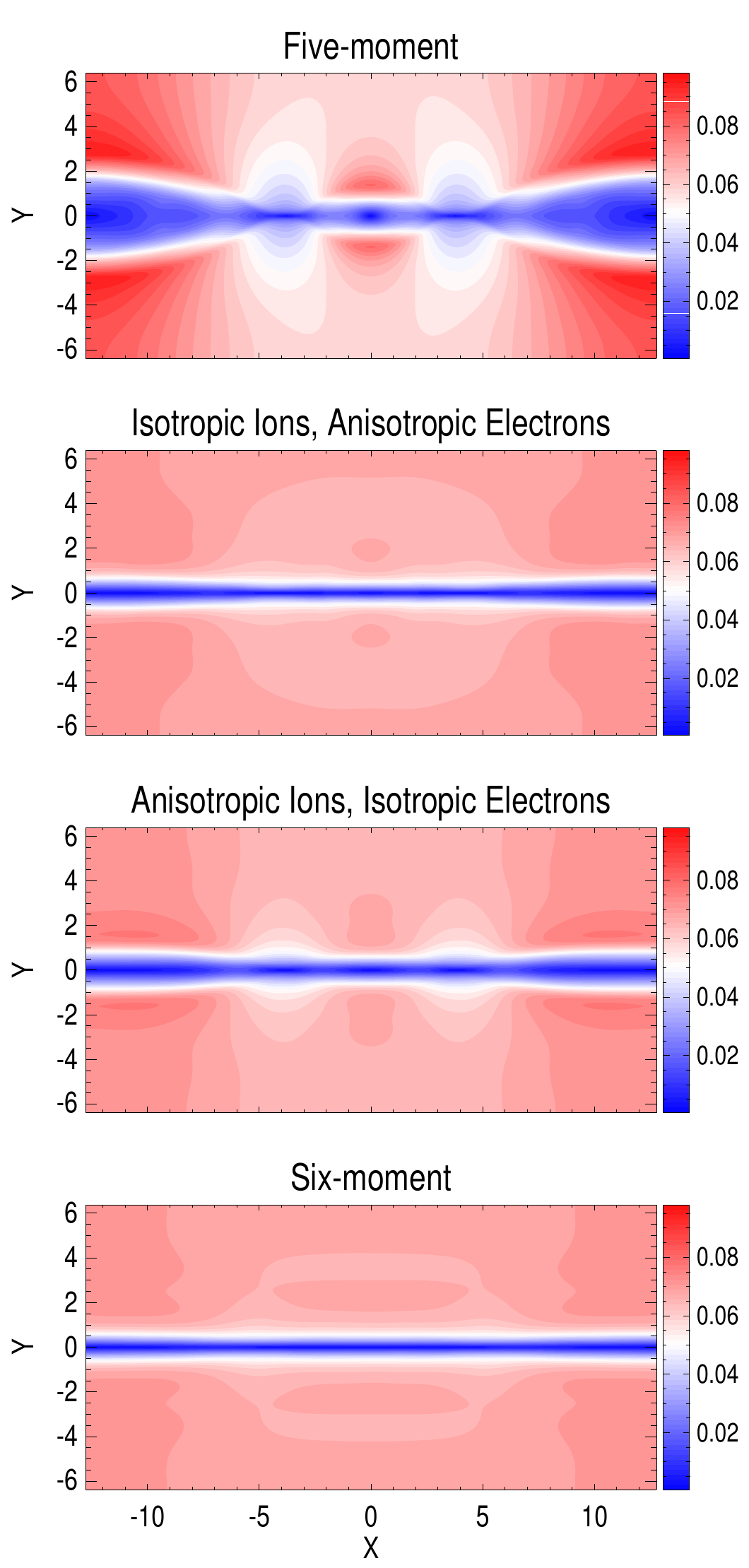}
\caption{The magnetic field magnitude for the four cases at $t = 495.01$.}
\label{fig:gem_t495_B}
\end{figure}

\begin{figure}[h!]
\center
\includegraphics[width=0.65\linewidth]{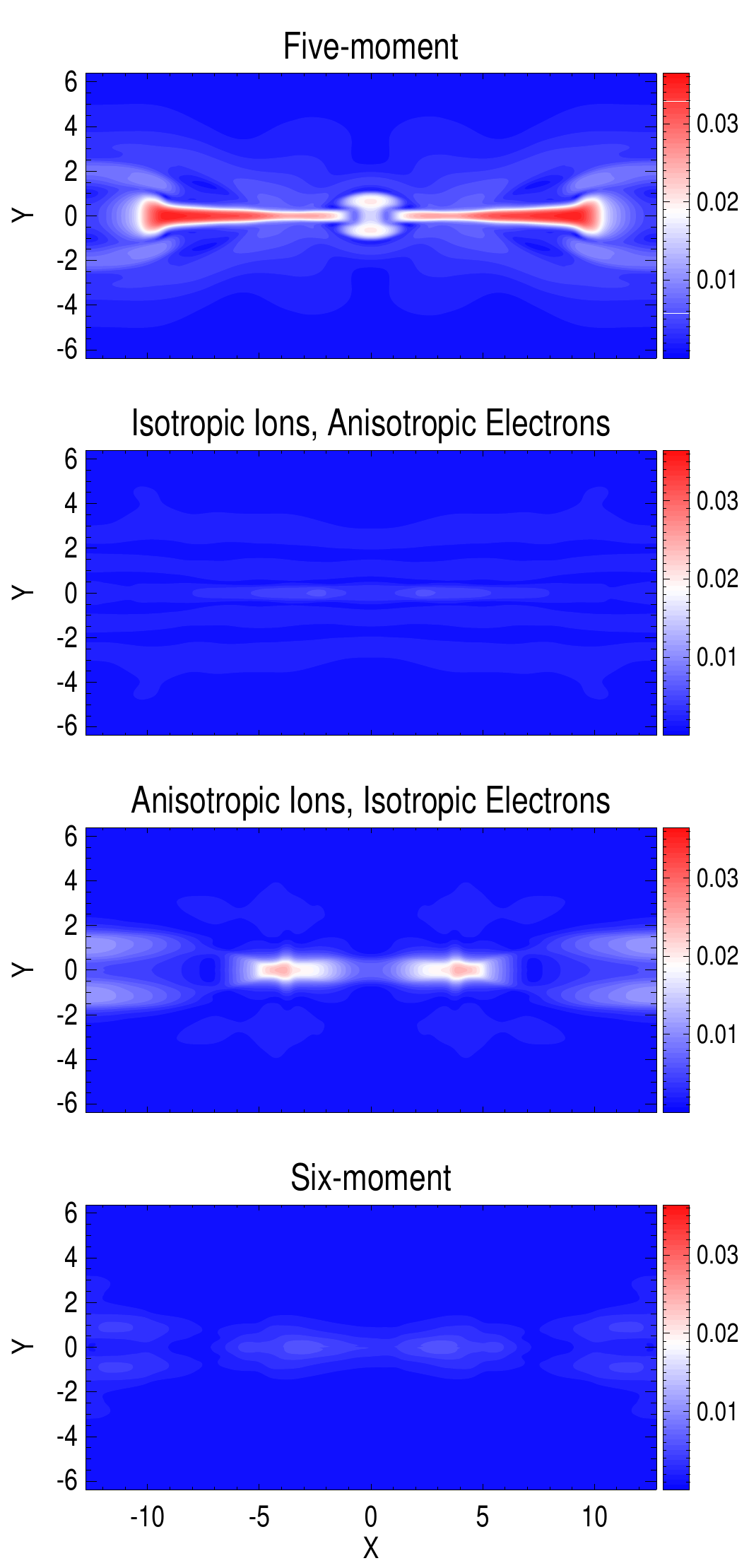}
\caption{The ion velocity magnitude for the four cases at $t = 495.01$.}
\label{fig:gem_t495_ui}
\end{figure}

\begin{figure}[h!]
\center
\includegraphics[width=0.65\linewidth]{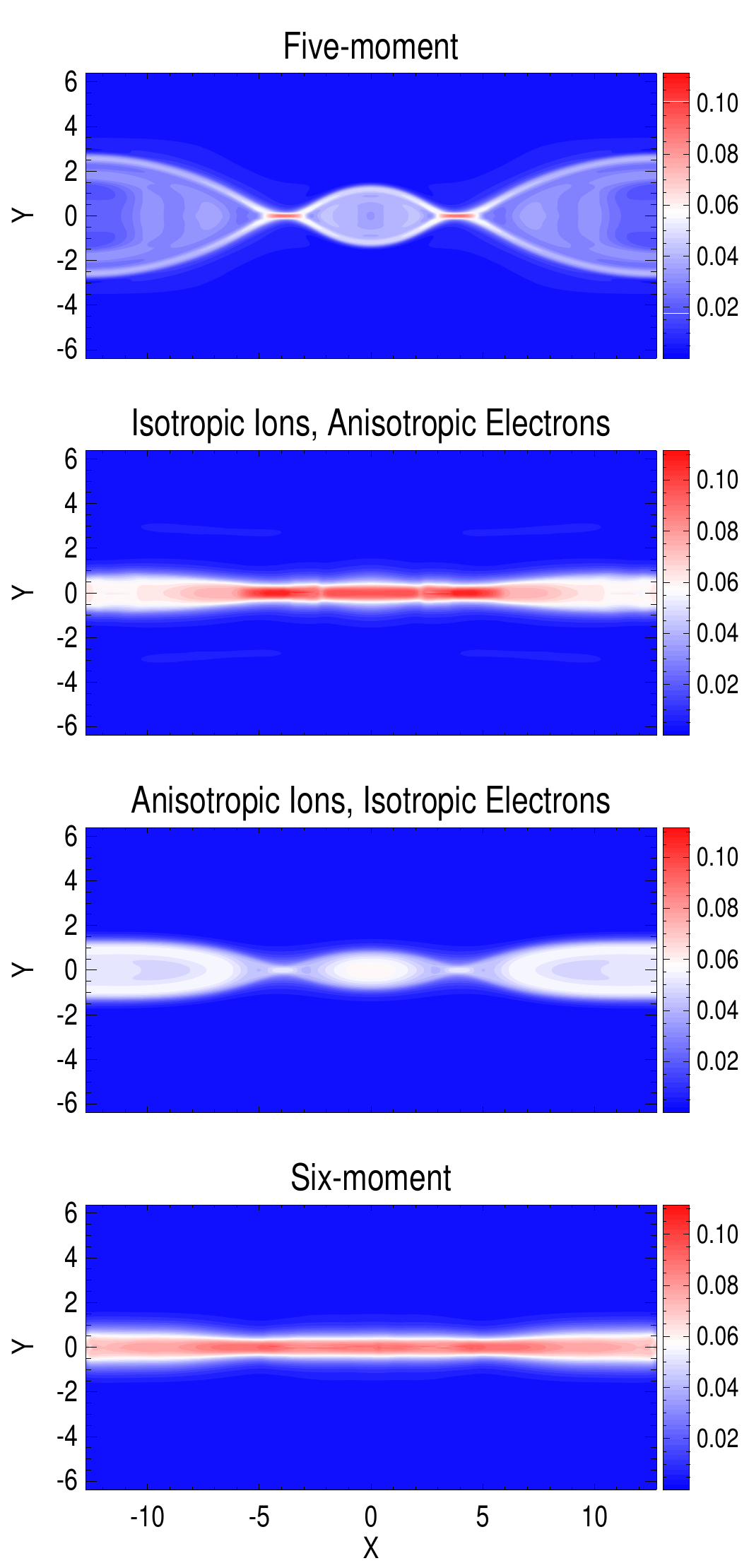}
\caption{The electron velocity magnitude for the four cases at $t = 495.01$.}
\label{fig:gem_t495_ue}
\end{figure}

\begin{figure}[h!]
\center
\includegraphics[width=0.7\linewidth]{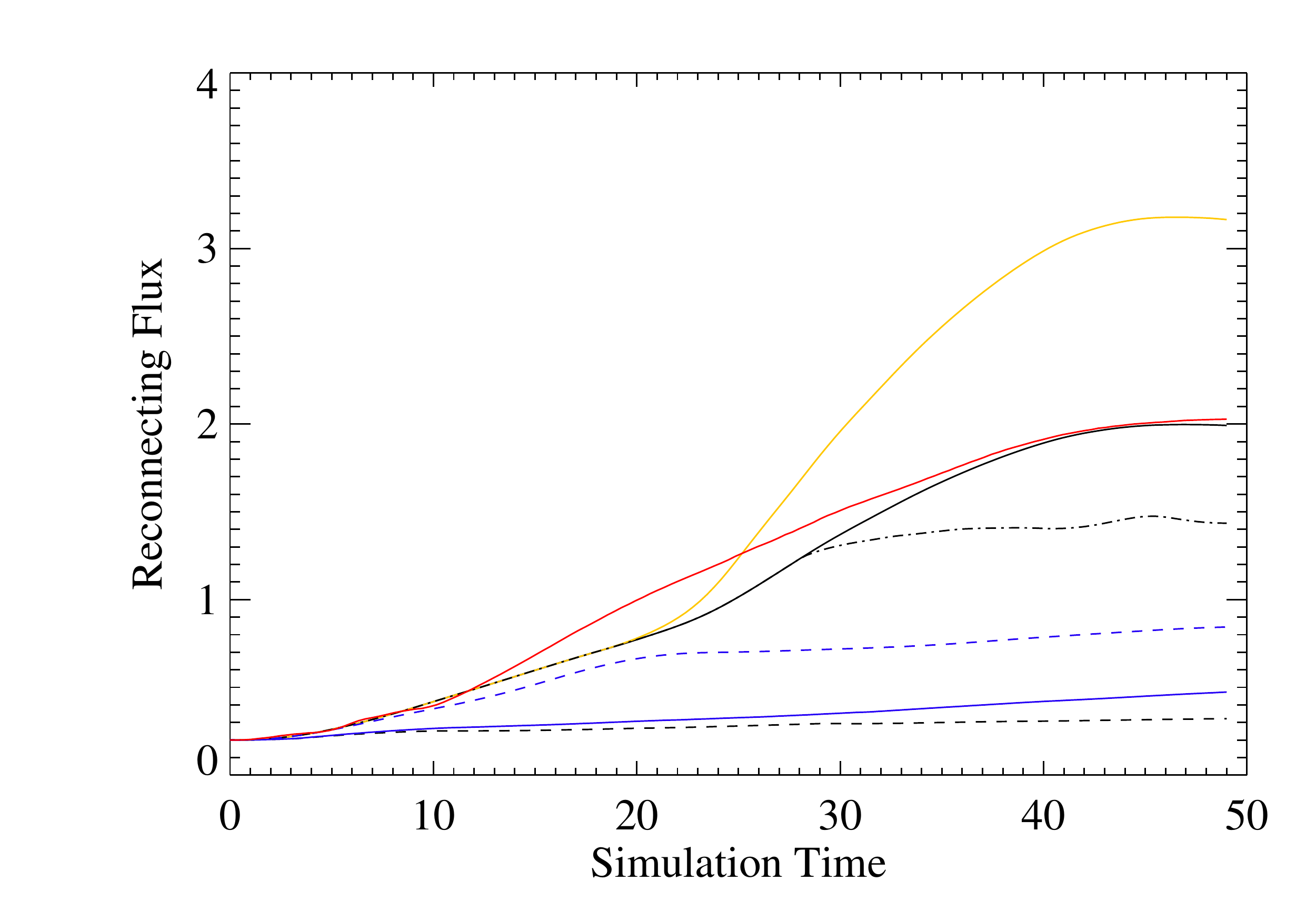}
\caption{The reconnected magnetic flux versus time. The simulation time is normalized to $\Omega_{ci}$ so that the
reconnected flux can be compared with Birn et al. \cite{Birn_2001}.
The yellow solid line shows the original reconnected flux obtained
from the approximate five-moment simulation while the black solid line shows the new reconnected flux from the same simulation.
The black dashed line is obtained from the pure six-moment simulation while the blue dashed/solid lines are from the isotropic ions/electrons cases, respectively.
The red solid line shows the reconnected flux from a PIC simulation from Chen et al. \cite{Chen_2018}. The black dashed dotted line
shows a simulation starting with $\tau = 1 \times 10^{-5}$ for both ions and electrons initially then turn off the relaxation constrains 
at $t=400$ (28.0 when normalized to $\Omega_{ci}$).}
\label{fig:gem_flux}
\end{figure}

In order to understand how pressure anisotropy affects the magnetic reconnection, we investigate four different scenarios: 1. the relaxation time $\tau = 1 \times 10^{-5}$ for both ions and electrons, 
which is an approximate five-moment simulation; 2. $\tau = 1 \times 10^{-5}$ for ions and no relaxation constrains for electrons, 
which is isotropic ions and anisotropic electrons; 3. $\tau = 1 \times 10^{-5}$ for electrons and no relaxation constrains for ions,
which is anisotropic ions and isotropic electrons; and 4. no relaxation for either ions or electrons, which is a six-moment
simulation.

Figures \ref{fig:gem_t495_B}, \ref{fig:gem_t495_ue} and \ref{fig:gem_t495_ui} show the magnetic field, 
ion velocity  and electron velocity magnitudes at $t = 495.01$ for the four cases, respectively.
The approximate five-moment simulation shows the same features as previous five-moment simulations performed in the literature \cite{Hakim_2006, Hakim_2008, Wang_2015}. The most surprising finding is that a pure six-moment simulation provides a completely different solution than
the five-moment simulation, despite the fact that the pressure anisotropy 
remains within 10\% of being isotropic for both ions and electrons during the whole simulation. 
Wang et al. \cite{Wang_2015} showed that with the full pressure tensor, their ten-moment
simulation can represent the magnetic reconnection process reasonably well compared to a PIC simulation.
This suggests that both the isotropic (five-moment) and full pressure tensor (ten-moment) equations give fast reconnection, but the anisotropic pressure (six-moment) does not.  From these four cases, we also find that
the isotropic electrons case is closer to the approximate five-moment case, which implies that the ion
pressure anisotropy has more impact than the electron anisotropy.

The reconnected magnetic flux $F = \int_{0}^{L_x/2} | B_y| dx$ is typically used to measure how fast
the magnetic reconnection occurs. In the approximate five-moment simulation, a magnetic island exists in the middle of the domain, so simply using this formula would provide a larger magnetic flux than the simulations
that do not have a magnetic island. The reconnected magnetic flux formula is modified to $F = \int_{0}^{L_x/2} \max(0, B_y) dx$, where only $B_y > 0$ is taken into account, so that the magnetic island does not contribute to the reconnected flux.

Figure \ref{fig:gem_flux} shows the reconnected flux for all cases. We find that the original reconnected flux formula provides a faster reconnection rate for the approximate five-moment simulation than the PIC simulation while the revised formula suggests a different result, which can be explained by the fact that the magnetic island in the five-moment simulation contributes to the reconnected flux in the original formula while it would not in the revised formula and no magnetic islands exist in the PIC simulation. 
We arrive at the conclusion that the revised formula provides a better 
description of the reconnected flux if magnetic islands exist. Figure \ref{fig:gem_flux} also shows that the pure six-moment simulation provides a very low reconnection
rate, as expected from Figure\,\ref{fig:gem_t495_B}\,-\,\ref{fig:gem_t495_ui}. Overall, the approximate five-moment simulation
has the closest reconnection rate to the PIC simulation. The isotropic electrons with anisotropic ions case has a faster reconnection rate than
the anisotropic electrons with isotropic electrons case, but they are both slower than the approximate five-moment simulation.
We also checked that if we start from the approximate five-moment simulation, then turn off the relaxation towards isotropy for both ions and electrons, then the already ongoing reconnection gets suppressed due to the developing
pressure anisotropies. This suggests that the six-moment equations produce slow reconnection even if the simulation is started from a fast reconnection scenario.

\section{Conclusions}
\label{sec:conclusion}

In this manuscript, we have developed a new model, the six-moment multi-fluid plasma model, to simulate
both ions and electrons with pressure anisotropy in a system when they can be described by the fluid equations. The new six-moment model solves for the full set of the electron continuity, momentum and pressure equations, as well as the exact
Maxwell equations. The six-moment model can simulate the light wave, Langmuir wave and
MHD waves accurately if the grid is fine enough to resolve the corresponding wave length. 

We use a steady-state conserving point-implicit time discretization for the stiff source terms  associated with the Lorentz force terms in the momentum equations and the $c^2 \mu_0 \mathbf{j}$ term in the Maxwell equations. The point-implicit time integration is combined with a spatial discretization based on a Lax-Friedrichs type numerical flux employing the fast magnetosonic speed for the plasma and magnetic field variables and the light speed to the electric field related variables. Using the magnetosonic speed instead of the light speed for the majority of the variables greatly reduces the numerical dissipation. Our numerical tests show that this approach is sufficient to maintain stability in most circumstances. Our implementation provides an option to use the speed of light for the Lax-Friedrichs flux for an arbitrary subset of variables if needed. 

A surprising discovery is that the six-moment model cannot provide a good description of the magnetic reconnection process. However, our goal is not to use the six-moment model to study the reconnection process. We plan to couple the six-moment model with an embedded PIC code \cite{Markidis_2010, Chen_2018}, similar as Daldorff et al. \cite{Daldorff_2014}. In such an approach, the magnetic reconnection region will be simulated with the PIC code and other regions will be simulated with the six-moment model. We expect that the six-moment equations are able to describe the vast majority of the plasma system outside magnetic reconnection regions where off-diagonal elements of the pressure tensor are negligible and with such an approach, the size of the PIC domain can be reduced.

An additional feature of the six-moment (also five- and ten-moment) equations is that it allows the use of multiple electron populations. The densities and velocities of multiple electron fluids cannot be approximated from charge neutrality and the current density, so an MHD approximation is not possible. This means that the six-moment equations can be applied to plasmas with thermal and super-thermal populations, or counter streaming populations, or populations of different origins (for example solar wind and ionospheric).

Finally, the six-moment equations provide a reasonable description of typical collisionless plasma conditions where the random motions along the field lines and the gyration perpendicular to the field naturally result in independent parallel and perpendicular pressures, but the off-diagonal terms are usually negligible in the vast majority of the simulation domain. The six-moment equations are only moderately more complicated than the 5-moment equations, but much simpler than the ten-moment equations, which results in lower computational cost and simpler implementation in comparison with the latter.

\section*{Acknowledgements}

This work was supported by the INSPIRE NSF grant PHY-1513379, the NSF
PREEVENTS grant 1663800, the NSF strategic 
capability grant AGS-1322543, the NASA grant NNX14AE75G, 
the US Rosetta Project with the JPL subcontract 1266313 under the Rosetta NASA grant NMO710889.

The authors would like to acknowledge the following high-performance computing resources:  
the Blue Waters super computer by the NSF PRAC grant ACI-1640510,
the Pleiades computer by NASA 
High-End Computing (HEC) Program through the NASA Advanced Supercomputing (NAS)
Division at Ames Research Center,
and Yellowstone (ark:/85065/d7wd3xhc) and Cheyenne 
(doi:10.5065/D6RX99HX) provided by NCAR's Computational and
Information Systems Laboratory, sponsored by the National Science Foundation. 

The six-moment solver is built within BATS-R-US and publicly available through the 
csem.engin.umich.edu/tools/swmf website after registration.

\newpage

\appendix
\section{Hyperbolic/parabolic cleaning for the Maxwell equations}

In this appendix, we briefly derive how the hyperbolic/parabolic cleaning works for the Maxwell equations.
We start from the modified Maxwell equations:

\begin{subequations}
\label{eqn:maxwell_appendix}
\begin{eqnarray}
\label{eqn:dB_appendix}
\frac{\partial \mathbf{B}}{\partial t} + \nabla \times \mathbf{E} + c_B \nabla \psi &=& 0							\\
\label{eqn:dE_appendix}
\frac{\partial \mathbf{E}}{\partial t} - c^2 \nabla \times \mathbf{B} + c_E \nabla \phi &=&
     - c^2 \mu_0 \mathbf{j}	\\
\label{eqn:divB_appendix}
\frac{\partial \psi}{\partial t} + c_B \nabla \cdot \mathbf{B} &=& -d_B \psi									\\
\label{eqn:divE_appendix}
\frac{\partial \phi}{\partial t} + c_E \nabla \cdot \mathbf{E} &=& \frac{c_E}{\varepsilon_0} {\rho_c} - d_E \phi
\end{eqnarray}
\end{subequations}

Taking ($\nabla \cdot$) of Equation (\ref{eqn:dB_appendix}) gives 
\begin{equation}
\label{eqn:B_step1}
\frac{\partial (\nabla \cdot \mathbf{B})}{\partial t} = - c_B \nabla^2 \psi
\end{equation}

With the expression $\nabla \cdot \mathbf{B} = -  \frac{1}{c_B} \frac{\partial \psi}{\partial t} - \frac{d_B}{c_B} \psi$
obtained from Equation (\ref{eqn:divB_appendix}), Equation (\ref{eqn:B_step1}) can be written as
\begin{equation}
\label{eqn:B_step2}
\frac{\partial^2 \psi}{\partial t^2} + d_B \frac{\partial \psi}{\partial t} = c_B^2 \nabla^2 \psi
\end{equation}
which is the damped wave equation, so the hyperbolic/parabolic cleaning variable $\psi$ 
propagates isotropically with speed $c_B$ and decays at a rate $d_B$.

Taking $(\frac{1}{c_B}\frac{\partial}{\partial t} + \frac{d_B}{c_B})$ of Equation\,(\ref{eqn:B_step1}) gives

\begin{equation}
\label{eqn:B_step3}
\left(\frac{1}{c_B}\frac{\partial}{\partial t} + \frac{d_B}{c_B}\right) \frac{\partial}{\partial t} (\nabla \cdot \mathbf{B}) = 
      - c_B \left(\frac{1}{c_B}\frac{\partial}{\partial t} + \frac{d_B}{c_B}\right) \nabla^2 \psi
\end{equation}
Taking the Laplace operator of the relationship $\nabla \cdot \mathbf{B} = -  \frac{1}{c_B} \frac{\partial \psi}{\partial t} - \frac{d_B}{c_B} \psi$ from Equation\,(\ref{eqn:divB_appendix}) gives
\begin{equation}
\nabla^2 (\nabla \cdot \mathbf{B}) = - \left(\frac{1}{c_B} \frac{\partial }{\partial t} + \frac{d_B}{c_B}\right)  \nabla^2 \psi
\end{equation}
so that Equation\,(\ref{eqn:B_step3}) becomes:

\begin{equation}
\label{eqn:B_step4}
\frac{\partial^2 (\nabla \cdot \mathbf{B})}{\partial t^2} + d_B \frac{\partial (\nabla \cdot \mathbf{B})}{\partial t} = c_B^2 \nabla^2 (\nabla \cdot \mathbf{B})
\end{equation}
which is the same damped wave equation as was obtained for  $\psi$ showing that $(\nabla \cdot \mathbf{B})$ will also propagate with speed  $c_B$ and decay at a rate $d_B$.

In a similar fashion, we take $\nabla \cdot$ of Equation (\ref{eqn:dE_appendix}) and obtain 
\begin{equation}
\label{eqn:E_step1}
\frac{\partial (\nabla \cdot \mathbf{E})}{\partial t} = -c^2 \mu_0 \nabla \cdot \mathbf{j} - c_E \nabla^2 \phi = - \frac{1}{\varepsilon_0} \nabla \cdot \mathbf{j} - c_E \nabla^2 \phi
\end{equation}

With the expression $\nabla \cdot \mathbf{E} = \frac{1}{\varepsilon_0} \rho_c - \frac{d_E}{c_E} \phi - \frac{1}{c_E} \frac{\partial \phi}{\partial t}$
obtained from Equation (\ref{eqn:divE_appendix}) and $\frac{\partial \rho_c}{\partial t} + \nabla \cdot \mathbf{j} = 0$
from the ion and electron continuity equations, 
Equation (\ref{eqn:E_step1}) can be written as
\begin{equation}
\label{eqn:E_step2}
\frac{\partial^2 \phi}{\partial t^2} + d_E \frac{\partial \phi}{\partial t} = c_E^2 \nabla^2 \phi
\end{equation}
which shows that the hyperbolic/parabolic cleaning variable $\phi$ for the electric field $\mathbf{E}$ 
propagates with $c_E$ and has a decay rate of $d_E$.

Taking $(\frac{1}{c_E}\frac{\partial}{\partial t} + \frac{d_E}{c_E})$ of Equation\,(\ref{eqn:E_step1}) gives

\begin{equation}
\label{eqn:E_step3}
(\frac{1}{c_E}\frac{\partial}{\partial t} + \frac{d_E}{c_E}) \frac{\partial}{\partial t} (\nabla \cdot \mathbf{E}) = - c_E (\frac{1}{c_E}\frac{\partial}{\partial t} + \frac{d_E}{c_E}) \nabla^2 \phi - \frac{1}{\varepsilon_0}(\frac{1}{c_E}\frac{\partial}{\partial t} + \frac{d_E}{c_E}) \nabla \cdot \mathbf{j}
\end{equation}

Taking the Laplace operator of $\nabla \cdot \mathbf{E} = -  \frac{1}{c_E} \frac{\partial \phi}{\partial t} - \frac{d_E}{c_E} \phi + \frac{1}{\varepsilon_0}\rho_c$ from Equation\,(\ref{eqn:divE_appendix}) gives $\nabla^2 (\nabla \cdot \mathbf{E}) = -  (\frac{1}{c_E} \frac{\partial }{\partial t} + \frac{d_E}{c_E}) \nabla^2 \phi + \frac{1}{\varepsilon_0} \nabla^2 \rho_c$, which can be substituted into Equation\,(\ref{eqn:E_step3}) and arrive at

\begin{equation}
\label{eqn:E_step4}
\frac{\partial^2}{\partial t^2} (\nabla \cdot \mathbf{E} - \frac{\rho_c}{\varepsilon_0}) + d_E \frac{\partial }{\partial t} (\nabla \cdot \mathbf{E}- \frac{\rho_c}{\varepsilon_0} ) = c_E^2 \nabla^2 (\nabla \cdot \mathbf{E} - \frac{\rho_c}{\varepsilon_0})
\end{equation}
which shows that $(\nabla \cdot \mathbf{E} - \frac{\rho_c}{\varepsilon_0})$ satisfies the damped wave equation and has the same behavior as $\phi$.

\newpage

\bibliographystyle{unsrt}
\bibliography{reference} % if your bibtex file is called example.bib

\end{document}